\documentclass[11pt]{aastex631}

\usepackage{graphicx}% Include figure files
\usepackage{dcolumn}% Align table columns on decimal point
\usepackage{bm}% bold math
\usepackage{longtable}
\usepackage{float,color}
\usepackage{subfigure}
\usepackage{overpic}
\usepackage{changepage}
\usepackage{mhchem}

\def\simge{\mathrel{\rlap{\raise 0.511ex
       \hbox{$>$}}{\lower 0.511ex \hbox{$\sim$}}}}
\def\simle{\mathrel{\rlap{\raise 0.511ex
        \hbox{$<$}}{\lower 0.511ex \hbox{$\sim$}}}}
   \shorttitle{Correlations Between the $M-R$ Relation and the EOS}
\shortauthors{Sun and Lattimer}
     
\begin{document}

\title{Correlations between the Neutron Star Mass-Radius Relation and the Equation of State of Dense Matter}

\correspondingauthor{B. Sun}
%\email{}

\author[0009-0005-9688-4369]{Boyang Sun}

\author[0000-0002-5907-4552]{James M. Lattimer}
\affiliation{Department of Physics \& Astronomy, Stony Brook University, Stony Brook, NY 11794 USA}

\begin{abstract}
We develop an analytic method of inverting the Tolman-Oppenheimer-Volkoff (TOV) relations to high accuracy. In principle, a specified $\mathcal{E}\mbox{-}P$ relation gives a unique $M\mbox{-}R$ relation, and vice-versa.  Our method is developed from the strong correlations that are shown to exist between the neutron star mass-radius curve and the equation of state (EOS) or pressure-energy density relation.  Selecting points that have masses equal to fixed fractions of the maximum mass, we find a semi-universal power-law relation between the central energy densities, pressures, sound speeds, chemical potentials and number densities of those stars, with the maximum mass and the radii of one or more fractional maximum mass points.  Root-mean-square fitting accuracies, for EOSs without large first-order phase transitions, are typically 0.5\% for all quantities at all mass points. The method also works well, although less accurately, in reconstructing the EOS of hybrid stars with first-order phase transitions. These results permit, in effect, an analytic method of inverting an arbitrary mass-radius curve to yield its underlying EOS.  We discuss applications of this inversion technique to the inference of the dense matter EOS from measurements of neutron star masses and radii as a possible alternative to traditional Bayesian approaches.
\end{abstract}

%% Keywords should appear after the \end{abstract} command. 
%% The AAS Journals now uses Unified Astronomy Thesaurus concepts:
%% https://astrothesaurus.org
%% You will be asked to selected these concepts during the submission process
%% but this old "keyword" functionality is maintained in case authors want
%% to include these concepts in their preprints.

\section{\label{sec:level1}Introduction}

Neutron stars provide a window into the equation of state of dense matter.  The standard Tolman-Oppenheimer-Volkoff (TOV) equations of general relativity \citep{Tolman34,Oppenheimer39}, combined with a specific energy density-pressure ($\mathcal{E}{\mbox-}P$) relation, or equation of state (EOS), generate a unique mass-radius ($M{\mbox-}R$) curve, where each point on the curve corresponds to a specific central density $\mathcal{E}_c$ and pressure $P_c$. The pressure and mass are integrated from the star's center, where the mass internal to the radius vanishes, to the surface, where the pressure vanishes.  As $P_c$ is increased, the stellar mass $M$ increases until a maximum mass $M_{max}$ is reached at the central pressure $P_{max}$, which is a general characteristic of the $M{\mbox-}R$ curve.  For higher values of $P_c$, the configuration becomes dynamically unstable.  At this mass, the energy density and pressure have their maximum possible values, $\mathcal{E}_{max}$ and $P_{max}$, respectively, for stable configurations.  The radius and sound speed have the values $R_{max}$ and $c_{s,max}/c=\sqrt{(\partial P/\partial\mathcal{E})_{max}}$.  However, unlike for $\mathcal{E}$ and $P$, $c_s$ may not be maximized at the star's center and  $R_{max}$ is smaller than its maximum possible value for that EOS.   Being the solution of a pair of first-order differential equations, the $M{\mbox-}R$ curve has a one-to-one correspondence with its associated EOS (i.e., the $\mathcal{E}{\mbox-}P$ relation), so knowledge of the $M{\mbox-}R$ curve can be used, in principle, to reconstruct this EOS. 

Observations of some neutron stars have provided simultaneous mass and radius estimates, but with typical uncertainties of several percent.   As these measurements become more abundant and accurate, methods to invert the structure equations will become more and more important.  Currently, a favored method of converting $M{\mbox-}R$ information, with their uncertainties, into information about the $\mathcal{E}{\mbox-}P$ relation involves Bayesian methodologies (see, e.g., \cite{Grinstead1997}). The existing schemes are mathematically complex, and generally involve the generation of an $M{\mbox-}R$ model space generated from a family of EOSs generated by varying parameters of a parameterized EOS.  This prior probability has an inherent uncertainty associated with the choice of parameterization as well as the probability distributions assumed for their parameters.  The parameters of these EOSs are varied between limits imposed by causality, hydrodynamic stability, and the necessity that each possible EOS should support a minimum maximum mass of about $2M_\odot$ resulting from pulsar timing measurements of neutron stars in binary systems. 

We call the model prior probability ${\cal P}({\cal M})$.  Each set of model parameters generates a particular model ${\cal M}_i$ that can generate data $P({\cal M}_i|{\cal D})$ in the form of an $M{\mbox-}R$ curve, for example.  Another input is the prior probability associated with the observed data, usually probability distributions in $M{\mbox-}R$ space, which we call ${\cal P}({\cal D})$. For many non-overlapping models ${\cal M}_j$ which exhaust the total model space $\cal M$, the prior data probability is equivalent to 
\begin{equation}
    {\cal P}({\cal D})=\sum_j{\cal P}({\cal D}|{\cal M}_j){\cal P}({\cal M}_j)=
    \int{\cal P}({\cal D}|{\cal M}){\cal P}({\cal M})d^N{\cal M},
 \label{eq:b1}   \end{equation}
    where $N$ is the number of possible models (i.e., number of parameter sets).
It is now assumed
that ${\cal P}({\cal D}|{\cal M})$, the conditional probability of the data given the model, is proportional to the product over the probability
distributions of the observed data ${\cal D}_i$ evaluated at the masses $M_i$ which are chosen in the model and evaluated at the radii which are determined from the model,
\begin{equation}
    {\cal P}({\cal D}|{\cal M})\propto\Pi_i{\cal D}_i.
\end{equation} 
Furthermore, if one chooses the parameters uniformly, which means all models have equal likelihood, ${\cal P}({\cal M})={\cal P}({\cal M}_j)$ and those terms cancel from Eq. (\ref{eq:b1}).
The posterior probability distribution for the model parameters is then computed using Bayes’ theorem,
\begin{equation}
    {\cal P}({\cal M}|{\cal D})={{\cal P}({\cal D}|{\cal M}){\cal P}({\cal M})\over{\cal P}({\cal D})}={\Pi_i{\cal D}_i\over\sum_j\Pi_i{\cal D}_i({\cal M}_j)}.
\end{equation}

This approach invariably has systematic uncertainties stemming from both prior choices of EOS parameterizations as well as the chosen distribution of their parameters (even if uniformly chosen, should these be in normal space or logarithmic space?).  These systematic uncertainties are not easily quantifiable, but in practice, by comparing results from different studies, we will show they are of roughly the same size as the observational uncertainties.

Other methodologies free of these prior model uncertainties might be possible if a sufficiently rapid and accurate technique of directly inverting the TOV equations existed.  A step in this direction was taken by \cite{Lindblom92} who proposed to integrate
instead from the star's surface to the center.  One assumes the EOS and $M{\mbox-}R$ relation is known up to some central density $\mathcal{E}_0$ and $P_0$ for which one has $M_0$ and $R_0$ as surface values.  Incrementally, one steps along the $M{\mbox-}R$ curve to a new value $M_1$ and $R_1$, and integrates towards the center.  As long as $P<P_0$, the EOS is known.  When the pressure reaches $P_0$ and density $\mathcal{E}_0$, one finds $r=r_1$ and $m(r_1)=m_1$.  An expansion of the TOV equations around $r=0$ then permits one to estimate the new central pressure and density $P_1$ and $\mathcal{E}_1$.  Starting from $M_2$ and $R_2$, one then infers $P_2$ and $\mathcal{E}_2$, etc., and successive steps can eventually extend to $M_{max}$ and $R_{max}$, generating the entire EOS in the process.   However, this scheme is difficult to implement as it is relatively unstable if too large steps are chosen, and errors accumulate too rapidly if too many steps are taken.  \cite{Lindblom92} suggests modifications to this simple strategy to improve convergence, but different $M{\mbox-}R$ curves may require different modifications.  Also, inversion speeds are compromised. A detailed overview of previous approaches for this problem is included in \S \ref{sec:prework}.

In this paper, we propose an analytic method to accurately invert the $M{\mbox-}R$ curve in order to obtain the $P\mbox{-}\mathcal{E}$ relation.  The inversion is accomplished through simple power-law fitting formulae determined from a sample of hundreds of published EOSs.  In \S \ref{sec:inv}, we will verify the existence of strong correlations among the quantities $M_{max},R_{max},\mathcal{E}_{max}$ and $P_{max}$, which have previously been pointed out by several recent publications, and further demonstrate that similar correlations exist for other points on the $M\mbox{-}R$ curve and their corresponding central EOS values.  We however note that these correlations are only accurate to order 5\% due to the non-uniqueness of $\mathcal{E}_c\mathcal{-}P_c$ values for a given $M\mbox{-}R$ point: two EOSs predicting the same $M\mbox{-}R$ values generally do not have the same $\mathcal{E}_c\mbox{-}P_c$ values.  In particular, concentrating on a specific fixed grid of fractional maximum mass points. and then, simultaneously using two of these points, we find correlations that accurately reproduce the underlying EOS at the centers of all these fractional mass stars.  The entire EOS up to $\mathcal{E}_{max}$ can then be determined by interpolation among these values and those of the core-crust interface. Furthermore, the central sound speeds, chemical potentials and baryon densities of these masses can similarly be determined through analogous correlations.

Unfortunately, prior lack of knowledge of $M_{max}$ and $R_{max}$ prevents this method from direct application to astronomical observations which yield $M\mbox{-}R$ uncertainty regions.  In a second approach  to more directly confront astronomical observations, in \S \ref{sec: ptop},  we demonstrate the existence of power-law correlations between an arbitrary $M\mbox{-}R$ points and their corresponding $P_c\mbox{-}\mathcal{E}_c$ values.  Their accuracy is limited to order 5\% due to the fact that a single $M-R$ point cannot uniquely specify $\mathcal{E}_c-P_c$ values.  The accuracy can be improved, however, by including more information about the observational $M\mbox{-}R$ region, and we focus on the inverse slope $dR/dM$ at the $M-R$ point that could be inferred from all observed $M\mbox{-}R$ uncertainy regions. We compare results of this approach with three published predictions based on traditional  Bayesian methods that assume a prior distribution taken from parametric EOSs, allowing us to estimate the systematic uncertainties of various approaches.

In \S \ref{sec:discuss}, we  demonstrate our technique for inverting an entire $M-R$ curve can also realistically reproduce the EOS of hybrid stars with first order phase transitions, although no hybrid EOSs were used to determine the parameters of our fitting formulae. We also discuss ways accuracy could be increased in the future,
although it is already far greater than needed given the current uncertainties of mass and radius estimates.  These include incorporating either sound speed and chemical potential information or an iterative procedure involving integration of the TOV equations using the approximately determined EOS.

\iffalse
This method cannot be applied successfully, currently, to observational data since the inferred masses and radii of individual stars has large uncertainties.  But having a method of directly transcribing an $M{\mbox-}R$ curve to an EOS suggests that a new type of Bayesian technique could be implemented, one in which one begins with parameterized $M{\mbox-}R$ curves instead of parameterized $\varepsilon{\mbox}P$ relations.  Furthermore, we will show that analytic relations can be found directly relating arbitrary mass and radius points to the corresponding central densities and pressures.  This has more relevance to converting mass-radius uncertainty regions into equation of state information by means of a parameter-free inversion.   This technique avoids the systematic uncertainties associated with prior choices of equation of state parametrizations and their parameters. 
\fi

\section{Previous Work Involving The Maximum Mass Point  \label{sec:prework}}

It is well-known that semi-universal (i.e., approximately EOS-independent) correlations exist relating aspects of the $M{\mbox-}R$ curve with properties of the EOS.  For example,  \cite{LP01} showed that the radii of typical (i.e., $M\approx1.4M_\odot$), are highly correlated with the neutron star matter pressure in the density range of $1-2n_s$, where $n_s$ is the nuclear saturation baryon density, about 0.16 fm$^{-3}$ (corresponding to an energy density $\mathcal{E}_s\simeq150$ MeV fm$^{-3}$).  This correlation has a several percent accuracy.

As a unique feature of the $M\mbox{-}R$ diagram, the maximum mass point has also received some attention \citep{Ofengeim2020,Cai2023,Ofengeim2023}.  The $M{\mbox-}R$ curve has a maximum mass $M_{max}$ point with a corresponding radius $R(M_{max})\equiv R_{max}$, central pressure $P_{max}$, energy density $\varepsilon_{max}$, and sound speed $c_{s,max}$.  \cite{Ofengeim2020} demonstrates that $\mathcal{E}_{max}, P_{max}$ and $c_{s,max}$ are correlated with $M_{max}$ and $R_{max}$ in an EOS-insensitive fashion using a set of 50 non-relativistic (Skyrme-like) and relativistic (RMF-like) nuclear interactions.  
He provided analytic fits for $M_{max}(\mathcal{E}_{max},P_{max})$, $R_{max}(\mathcal{E}_{max},P_{max})$ and $c_{s,max}(\mathcal{E}_{max},P_{max})$ as well as their inverses $\mathcal{E}_{max}(M_{max},R_{max})$, $P_{max}(M_{max},R_{max})$ and $c_{s,max}(M_{max},R_{max})$.

Utilizing a larger database of over 500 non-relativistic and relativistic interaction models tabulated by \cite{Sun2024}, we reexamined these fits.  We restricted our EOS dataset to the subset (316) that satisfied $M_{max}\ge2M_\odot$.  For every interaction, we assumed a common SLy4 \citep{chabanat1998skyrme} crustal EOS below the density 0.04 fm$^{-3}$ in the form of a piecewise polytrope (PP) fit \citep{Zhao2022}.  The uniform matter EOS from every interaction was used above the nuclear saturation density $n_s=0.16$ fm$^{-3}$.  A smooth interpolation in-between these two densities was provided by a cubic polynomial fit that guaranteed continuity of $\mathcal{E}, P$ and $c_s$ at the endpoints. The TOV equations were integrated using $\ln P$ as the independent variable with a surface boundary pressure $10^{-10}$ MeV fm$^{-3}$.  We verified that lowering the surface pressure or choosing an alternate crustal EOS did not significantly affect the structural properties of neutron stars.

We found  \cite{Ofengeim2020}'s relations fit $\mathcal{E}_{max}$, $P_{max}$ and $c_{s,max}$ to $M_{max}$ and $R_{max}$ data with root-mean-square (RMS) errors of about 12\%, 18\% and 4.5\%, respectively.  
\cite{Cai2023} improved these correlations, starting from a theoretical rationale for the existence of correlations among $\mathcal{E}_{max}, P_{max}, R_{max}$ and $M_{max}$ based on a truncated perturbative expansion (TPE) of the TOV equations \citep{Cai2023}.  Using the scale length
\begin{equation}
    Q=\sqrt{c^4\over4\pi G\mathcal{E}_c},
    \end{equation} 
    where $\mathcal{E}_c$ is the central energy density, the dimensionless radius, mass interior to this radius, energy density and pressure can be defined
    \begin{equation}
        \hat r={r\over Q},\qquad \hat m={Gm\over Qc^2}, \qquad \hat {\cal E}={{\cal E}\over{\cal E}_c},\qquad \hat  P={P\over{\cal E}_c}.
    \end{equation}
    The corresponding dimensionless TOV equations become
    \begin{equation}
        {d\hat P\over d\hat r}=-{(\hat  {\cal E}+\hat P)(\hat m+\hat r^3\hat P)\over\hat r(\hat r-2\hat m)},\qquad {d\hat m\over d\hat r}=\hat{\cal E}\hat r^2.
    \label{eq:TOV}\end{equation}
    Expanding the mass, energy density and pressure 
    in powers of the dimensionless radius $\hat r$, such that
    \begin{equation}
        \hat{\cal E}=\sum_ia_i\hat r^i,\qquad \hat P=\sum_ib_i\hat r^i,\qquad \hat m=\sum_ic_i\hat r^i,
    \end{equation}
  one finds the non-vanishing leading-order coefficients
  \begin{equation}
  a_0=1,\qquad a_2=-{1\over \hat R^2},\qquad b_0=\hat P_c,\qquad b_2=-{1+4\hat P_c+3\hat P_c^2\over6},\qquad c_3={a_0\over3},\qquad c_5={a_2\over5},
  \end{equation}
  where $\hat P_c$ is the dimensionless central pressure.  The expansions are truncated at second order and it is required that the pressure $\hat P$ and energy density $\hat{\cal E}$ vanish at the stellar surface where $\hat r=\hat R$ and $\hat m=\hat M$.  The coefficients $a_1=b_1=c_0=c_1=c_2=c_4=0$ vanish due to  symmetry, which requires that first derivatives of $\varepsilon, P$ and $m$ vanish at the origin.     
  In particular, for the maximum mass configuration
\begin{eqnarray}
\hat M_{max}&=&{GM_{max}\over c^2}\sqrt{4\pi G{\cal E}_{max}\over c^4}\simeq{2\hat R_{max}^3\over15},\cr   
\hat R_{max}&=&R_{max}\sqrt{4\pi G\mathcal{E}_{max}\over c^4}\simeq\sqrt{6\hat P_{max}\over1+4\hat P_{max}+3\hat P_{max}^2}\equiv\sqrt{6\phi_{max}},\cr
   c^2_{s,max}\over c^2&=&\left({dP\over d{\cal E}}\right)_{max}={b_2\over a_2}\simeq\hat P_{max},
\label{eq:caifit}\end{eqnarray}
which defines the function $\phi$.  We note that \cite{Cai2023} did not require the energy density to vanish at the surface, and thereby obtained $\hat M_{max}=\hat R_{max}^3/3$ instead. \cite{Cai2023} recognized that  Eq. ({\ref{eq:caifit}) itself was a poor fit to equation of state data, and therefore proposed the following linear fits based on the same scaling relations:
\begin{equation}
M_{max}=\alpha_M+\beta_M\left({\mathcal{E}_{max}\over{\rm GeV~fm}^3}\right)\nu_{max}^3, \qquad R_{max}=\alpha_R+\beta_R\nu_{max},
\label{eq:caifit1}\end{equation}
where
\begin{equation}
\nu_{max}=\sqrt{\phi_{max}{\rm~GeV~fm}^{-3}\over\mathcal{E}_{max}}.
\label{eq:nu}
\end{equation}
Their fits were claimed to be accurate to 0.39 km and $0.05M_\odot$, respectively.  This expansion cannot be carried out to higher order than second without the specification of equation of state information.

We refitted these TPE relations using the larger equation of state sample from \cite{Sun2024}, and found
$\alpha_R=0.9776$ km, $\beta_R=31.76$ km, $\alpha_M=-0.0519M_\odot$, and $\beta_M=53.65M_\odot$ with fitting accuracies of 0.23 km and $0.037M_\odot$, or 2.0\% and 1.6\%, respectively.  Note the original relations Eq. (\ref{eq:caifit}) would have implied that $\beta_R=19.0$ km and $\beta_M=10.3M_\odot$.

However, we are more interested in the inverse of this procedure, namely finding $\mathcal{E}_{max}$ and $P_{max}$ from $M_{max}$ and $R_{max}$. We note that Eqs. (\ref{eq:caifit1}) and (\ref{eq:nu}) imply $\nu_{max}$ scales with $R_{max}$ while $\phi_{max}$ scales with $M_{max}/R_{max}=\hat M_{max}/\hat R_{max}$ and $c_{s,max}$ scales with $\sqrt{M_{max}/R_{max}}$, so one can instead attempt to fit
\begin{equation}
\nu_{max}=a_\nu+b_\nu R_{max},\qquad\phi_{max}=a_\phi+b_\phi \left({GM_{max}\over R_{max}c^2}\right),\qquad
{c_{s,max}\over c}=\alpha_c+\beta_c\sqrt{GM_{max}\over R_{max}c^2}
%    {\mathcal{E}_{max}\over{\rm GeV~fm}^{-3}}={\beta_R^3\over \beta_M}{M_{max}-\alpha_M\over(R_{max}-\alpha_R)^3},
%    \qquad \phi_{max}={\beta_R\over \beta_M}{M_{max}-\alpha_M\over R_{max}-\alpha_R}
.\label{eq:caifit2}\end{equation}
%   With the above fitting parameters, we find that $\mathcal{E}_{max}$ is fit to an accuracy of 7.5\%.   
The proportionality relation for $c_{s,max}$ can be justified, somewhat arbitrarily, if one ignores the denominator term in the relation for $\hat R_{max}$ in Eq. (\ref{eq:caifit}) such that $\hat P_{max}\propto\hat R_{max}^2$; then, $c_s^2\propto (\hat M_{max}/\hat R_{max}^3)\hat R_{max}^2$ where the smallest integer exponent of $\hat M_{max}/\hat R_{max}$ was sought.  A referee kindly pointed out that \cite{Cai2023vs}, keeping higher order terms and requiring $dM/d\mathcal{E}=0$ at $r=0$ instead of $\mathcal{E}=0$ at $r=R$, found a more accurate relation such that $c_{s,max}\propto (M_{max}/R_{max})^2$.

Fitting these relations to the compilations of  \cite{Sun2024}, it is found that $a_\nu=-0.0229$, $b_\nu=0.0309$ km$^{-1}$, $a_\phi=0.0724$, $b_\phi=0.200$, $\alpha_c=-0.776$ and $\beta_c=5.61$, and $\nu_{max}$, $\phi_{max}$ and $c_{s,max}/c$ are fitted to the accuracies of 1.7\%, 1.1\% and 5.9\%, respectively. Scaling $c_{s,max}/c$ instead with $(M_{max}/R_{max})^2$ does not substantially change its fitted accuracy. $\mathcal{E}_{max}=(\phi_{max}/\nu_{max}^2)$ GeV fm$^{-3}$ is determined to 4.2\% accuracy, but
\begin{equation}
\hat P_{max}={1\over3\phi_{max}}\left[\left({1\over2}-2\phi_{max}\right)-\sqrt{\phi_{max}^2-2\phi_{max}+{1\over4}}\right]
\label{eq:pmax}\end{equation}
cannot always be determined from this sample of equations of state since Eq. (\ref{eq:pmax}) has no real roots when $\phi_{max}>1-\sqrt{3}/2$ or $\hat P_{max}>1/\sqrt{3}$, which is true of a number of equations of state in the tabulation.  Nevertheless, further restricting the equation of state pool such that the predicted values of $\phi_{max}\le1-\sqrt{3}/2$, which removes 50 equations of state and reduces the largest maximum mass of the sample to $3.16M_\odot$ (previously, it was $3.25M_\odot$), yields fits to $\hat P_{max}$ and $P_{max}$ accurate to 6.1\% and 7.1\%, respectively.  It is important to note that predicting the EOS from $M{\mbox-}R$ data is less precise than the inverse.

The TPE approach giving Eq. (\ref{eq:caifit}) suggests there exists a roughly power-law relationship among $\mathcal{E}_{max}$, $P_{max}$, $c_{s,max}$, $M_{max}$ and $R_{max}$.  Based on this, \cite{Ofengeim2023} introduced fits that closely resemble power-law relations, and, based on a set of 160 nuclear interactions, found accuracies for $M_{max}(\mathcal{E}_{max},P_{max})$, $R_{max}(\mathcal{E}_{max},P_{max})$ and $c_{s,max}(\mathcal{E}_{max},P_{max})$ of 0.86\%, 2.0\% and 4.9\%, respectively, which would represent improvements to not only the earlier fits of \cite{Ofengeim2020}, but also the fits inspired by \cite{Cai2023vs}.  However, we have not been able to completely reproduce the results of \cite{Ofengeim2023}.  Fitting to the set of equations of state in the tabulation of \cite{Sun2024}, we find the  formulae from \cite{Ofengeim2023} yield accuracies of  9.1\%, 2.3\%, and 11.1\%, respectively.  
This discrepancy could be due to the different suite of EOSs considered here, as well as our neglect of their non-linear terms.
%This discrepancy could be due to typos in \cite{Ofengeim2023}.

Noting that the fits of \cite{Ofengeim2023} are power-law fits if one ignores certain relatively small constant terms, and that they apparently are superior the TPE fits, 
%and that they are also the apparent inverse of the approach discussed in Ref. \cite{Ofengeim2023}, 
we pursue a power-law approach further.  First we compare directly to \cite{Ofengeim2023} by seeking fits to the quantities $G\in[M_{max},R_{max}]$ from $\mathcal{E}_{max}$ and $P_{max}$ from the purely power-law relations
\begin{eqnarray}
    G&=&a_G\left({P_{max}\over{\rm MeV~fm}^{-3}}\right)^{b_G}\left({\mathcal{E}_{max}\over{\rm GeV~fm}^{-3}}\right)^{c_G}
%    ,\cr R_{max}&=&a_R\left({P_{max}\over{\rm MeV~fm}^{-3}}\right)^{b_R}\left({\mathcal{E}_{max}\over{\rm GeV~fm}^{-3}}\right)^{c_R}
\label{eq:power}\end{eqnarray}
and determining the fitting parameters by minimizing $\chi^2$ with respect to them, where
  \begin{equation}
\chi_G^2=N^{-1}\sum_i^N\left[\ln\left({G_i\over a_G}\right)-b_G\ln\left({P_{max,i}\over{\rm MeV~fm}^{-3}}\right)-c_G\ln\left({\mathcal{E}_{max,i}\over{\rm GeV~fm}^{-3}}\right)\right]^2,
  \label{eq:G}\end{equation} 
  with $i$ running over the $N=316$ equations of state from \cite{Sun2024} that have $M_{max}\ge2.0M_\odot$.    We define the relative (logarithmic) error for an individual equation of state $i$ by
    \begin{equation}
\delta G_i={a_{G}\over G_i}\left({P_{max,i}\over{\rm MeV~fm}^{-3}}\right)^{b_{G}}\left({\mathcal{E}_{max,i}\over{{\rm GeV~fm}^{-3}}}\right)^{c_{G}}-1.
\label{eq:rmsi}
\end{equation}
It is found that $M_{max}$ and $R_{max}$ can be fit, with root mean square (RMS) relative errors defined by
\begin{equation}
<\delta G>=\sqrt{N^{-1}\sum_i^N(\delta G_i)^2},
      \label{eq:rms}
  \end{equation}
  of 0.61\% and 0.79\%,  respectively. 
(We note that in the rest of this paper, accuracy means the RMS error defined in this way.)  The associated parameters, $a_M=1.136M_\odot$, $b_M=0.278$, $c_M=-0.802$, $a_R=2.50$ km, $b_R=-0.0325$ and $c_R=-0.433$.
have relatively similar effective exponents implied by the formulae from \cite{Ofengeim2023} after removing their small non-linear term.  Interestingly, we find these fits to be slightly superior to those of \cite{Ofengeim2023}.  In comparison, it seems curious that the TPE approach [Eq. (\ref{eq:caifit1})] predicts qualitatively different exponents, $b_M\sim3/2$ %, $c_M\sim-2$, 
 and $b_R\sim1/2$ % and $c_R\sim-1$ 
 if we ignore the constant terms $\alpha_M$ and $\alpha_R$.

  \begin{table}
  \centering
      \begin{tabular}{c|cccc|cccc} \toprule \\[-9pt]
    $f=M_f/M_{max}$ & $a_{\mathcal{E}f}$ & $b_{\mathcal{E}f}$ & $c_{\mathcal{E}f}$
    & $<\delta\mathcal{E}_f>$ & $a_{\nu f}$ & $b_{\nu f}$ & $<\delta\mathcal{E}_{TPE,f}>$ & $<\delta\nu_{TPE,f}>$
    \\ \hline \\[-10pt]
  1  &  1.948   &  -0.2675    &  -1.861   &  0.0153   & -0.02054   &  0.03048
 &    0.0434  &   0.0216
  \\
 0.95& 1.322  & -0.1255 & -2.118  & 0.0262 
 & -0.05213 & 0.03656 & 0.0587 & 0.0262
 \\
 0.9  & 1.138  & -0.0588 & -2.213  & 0.0302
 &-0.06389 & 0.03841 & 0.0634 & 0.0305
 \\
 0.85  & 1.014  & -0.009629 & -2.275  & 0.0332
 & -0.07068 & 0.03948 & 0.0666 & 0.0318
 \\ 
4/5 & 0.9154   & 0.03423 & -2.328   & 0.0359 
& -0.07408 & 0.04009 & 0.0691 & 0.0326
\\
3/4  & 0.8352  & 0.06797 & -2.362 & 0.0383  & -0.07557 & 0.04042 &0.0713 & 0.0333  \\
2/3 & 0.7221  & 0.1166& -2.409   & 0.0428
& -0.07485 & 0.04056 & 0.0747 & 0.0342
\\
3/5 & 0.6465  & 0.1462 & -2.442   & 0.0469
& -0.07259 & 0.04041 & 0.0778 & 0.0350
\\ 
1/2 & 0.5533  & 0.1623 & -2.466  & 0.0551
& -0.06803 & 0.03990 & 0.0843 & 0.0371
\\
2/5 & 0.4811 & 0.1229 &-2.431 & 0.0664 
& -0.06276 & 0.03904 & 0.0956 & 0.0418
\\
1/3  & 0.4412 & 0.0567 & -2.351 & 0.0758 
& -0.05286 & 0.03818 & 0.107 & 0.074
\\ \hline\hline\\[-9pt]
%\end{tabular}    
%       \begin{tabular}{c|cccc|ccc} \toprule
    $f=M_f/M_{max}$ & $a_{Pf}$ & $b_{Pf}$ & $c_{Pf}$ & $<\delta P_f>$ 
    & $a_{\phi f}$ & $b_{\phi f}$ & $<\delta P_{TPE,f}>$ & $<\delta\phi_{TPE,f}>$
    \\ \hline 
  1  & 0.1213 & 2.747 & -5.245 & 0.0454 & 0.07346 & 0.1978 & 0.0668 & 0.0216
  \\
 0.95  & 0.1035  & 2.229 & -4.661  & 0.0457& 0.03462 & 0.3232 & 0.0881 & 0.0101 
 \\
 0.9& 0.08867  & 2.075 & -4.492  & 0.0468 & 0.02351 & 0.3473 & 0.0825 & 0.0102
 \\
 0.85  & 0.07644  & 1.968 & -4.378  & 0.0485
 & 0.01681 & 0.3531 & 0.0821 & 0.0104
 \\ 
4/5 & 0.06564   & 1.891 & -4.290   & 0.0506
& 0.01242 & 0.3500 & 0.0691  & 0.0108
\\
3/4  & 0.05675 & 1.816  & -4.214 & 0.0528 & 0.009626  & 0.3396 & 0.0840 & 0.0115  \\
2/3 & 0.04404  & 1.714 & -4.099  & 0.0575
& 0.006813 & 0.3140 & 0.0869 & 0.0125
\\
3/5 & 0.03582  & 1.637 & -4.027  & 0.0625 
& 0.005730 & 0.2880 & 0.0904 & 0.0137
\\ 
1/2 & 0.02610  & 1.500 & -3.898  & 0.0726 & 0.005659 & 0.2421 & 0.0982 & 0.0157
\\
2/5 & 0.01882 & 1.302 & -3.686 & 0.0860 & 0.006764 & 0.1914 & 0.111 & 0.0173\\
1/3 & 0.01478 & 1.312 & -3.457 & 0.0960 & 0.007520 & 0.1576 & 0.123 & 0.0177\\
\hline\hline\\[-9pt]
%\end{tabular}    
%
%      \begin{tabular}{c|cccc|ccc} \toprule
    $f=M_f/M_{max}$ & $a_{cf}$ & $b_{cf}$ & $c_{cf}$ & $<\delta c_{s,f}>$
    & $\alpha_{cf}$ & $\beta_{cf}$& $<\delta c_{s,TPE,f}>$
    \\ \hline 
  1  & 0.2722 & 1.814 & -2.258 & 0.0660 & -0.7067 & 5.402 & 0.0818
  \\
 0.95  & 0.3192  & 1.592  & -1.825 & 0.0239 &-0.4279 &4.997 & 0.0330 
 \\
 0.9& 0.3246  &1.503  & -1.654 & 0.0195 &-0.3450 &4.133 & 0.0240
 \\
 0.85  & 0.3212  & 1.449 & -1.544  & 0.0175  &-0.2984 &3.878 & 0.0194
 \\ 
4/5 & 0.3147   & 1.408  & -1.461  & 0.0170 &-0.2644 &3.660 & 0.0177
\\
3/4  & 0.3073 & 1.371 & -1.399 & 0.0178   & -0.2346 & 3.450  & 0.0181   \\
2/3 & 0.2944  & 1.311  & -1.315 & 0.0202  &-0.1885 &3.101 & 0.0205
\\
3/5 & 0.2849  & 1.259  & -1.266 & 0.0227  &-0.1506 & 2.810 & 0.0231 
\\ 
1/2 & 0.2727  & 1.162  & -1.199 & 0.0269 & -0.08919 & 2.345& 0.0274 
\\
2/5 & 0.2612 & 1.039 & -1.119 & 0.0315 & -0.02633 & 1.866 & 0.0322\\
1/3 & 0.2531 & 0.946 & -1.063 & 0.0357 & -0.01238 & 1.556 & 0.0370 \\
\hline\hline\\[-9pt]
    $f=M_f/M_{max}$ & $a_{\hat Pf}$ & $b_{\hat Pf}$ & $c_{\hat Pf}$ & $<\delta\hat P_f>$
    &  & & $<\delta\hat P_{TPE,f}>$
    \\ \hline 
1 &0.06226 & 3.014 &-3.383& 0.0320 & && 0.0613\\
0.95&0.07827 &2.354 &-2.543 & 0.0213 && & 0.0421\\
0.9&0.07790 & 2.134 & -2.280 & 0.0190 & && 0.0310\\
0.85& 0.07534 &1.978&-2.102& 0.0183& & &0.0266\\
4/5& 0.07171 &1.857&-1.966& 0.0180 & & &0.0241\\
3/4& 0.06100&1.597 &-1.690 &0.0187 & & &0.0230\\
2/3&0.05772&1.648&-1.745&0.0177 & & &0.0218\\
3/5& 0.05540 & 1.491 & -1.585 & 0.0195 && & 0.0220\\
1/2 & 0.04717 & 1.337 & -1.432 & 0.0209 && &0.0227\\
2/5 & 0.03912 & 1.179 & -1.255 & 0.0222 &&& 0.0231\\
1/3 & 0.03351 & 1.075 & -1.106 & 0.0224 &&& 0.0226\\
\hline\hline
\end{tabular}    

\caption{The parameters and RMS errors $<\!\!\delta G\!\!>$ for the power-law fits in Eq. (\ref{eq:powfit1}) are in the center blocks and those for the TPE fits in Eqs. (\ref{eq:caifit2}) and (\ref{eq:pmaxf}) are in the right blocks.
% and the right section contains parameters for the TPE fits in Eq. (\ref{eq:caifit2f}). 
$a_{\mathcal{E}f}$ and $a_{Pf}$ have units of GeV fm$^{-3}$ and $b_{\nu f}$ has units of km$^{-1}$. }
     \label{tab:powfit}
\end{table}

Since we are primarily interested in the inverse problem, we now seek power-law fits for $G\in[\mathcal{E}_{max}, \, P_{max}, \, c_{s,max}/c, \, \nu_{TPE,max}, \, \phi_{TPE,max}, \, c_{sTPE,max}/c]$ in terms of $M_{max}$ and $R_{max}$,
 \begin{eqnarray}
     G&=&a_{G}\left({M_{max}\over M_\odot}\right)^{b_{G}}\left({R_{max}\over{10\rm~km}}\right)^{c_{G}}.
   \label{eq:powfit}\end{eqnarray}
%The coefficients and effective exponents are presented in the first rows ($f=1$) of each block of entries in Table \ref{tab:powfit}.
%$a_\mathcal{E}=1.907$ GeV fm$^{-3}$,  $b_\mathcal{E}=-0.250$, $c_\mathcal{E}=-1.864$, 
%$a_P=0.1109$ GeV fm$^{-3}$, $b_P=2.832$, $c_P=-5.331$, $a_c=0.2605$, $b_c=1.857$ and $c_c=-2.297$.  
These fits for maximum mass values of $\mathcal{E}, P$ and $c_s$ turn out to be accurate to 1.5\%, 4.5\%, and 3.6\%, respectively, in the power-law case,  an improvement over the RMS errors of 4.2\%, 7.1\% and 5.9\%, respectively, for the TPE approach.  We again see that predicting $(M_{max},R_{max})$ from the EOS is more accurate than the inverse situation.

Dimensionally, one naively expects that $b_\mathcal{E}$ and $c_\mathcal{E}$ would be approximately 1 and 3, respectively; however, the best-fit values are $b_\mathcal{E}\sim-0.27$ and $c_\mathcal{E}\sim-1.86$, respectively, which means the $\mathcal{E}_{max}$ fit differs from the expected behavior by an approximate factor $R/M$.  For the pressure, we find $b_P\sim2.74$ and $c_P=-5.25$ compared to the values of 2 and -4 expected on dimensional grounds; the differing behavior is approximately a factor of $M/R$.  Finally, the sound speed behavior is found to be approximately $(M/R)^2$ whereas, previously, we estimated a scaling $\sqrt{M/R}$.

This apparently anomalous behavior is, partly, explained by our omission of higher order terms in $\hat P_{max}$ that are considered by \cite{Cai2023vs}. When included, as a referee kindly pointed out, the TPE method gives good estimates for some of the relevant scalings. The factor $6\hat P_{max}/(1+4\hat P_{max}+3\hat P_{max}^2)$ can, for example, be approximated by $\hat P_{max}^{~0.4}$, so that $R_{max}$ scales as $P_{max}^{~0.2} \mathcal{E}_{max}^{-0.7}$. Similarly, one can establish that $M_{max}$ scales as $ P_{max}^{~0.7} \mathcal{E}_{max}^{-1.2}$, so that $P_{max}$ scales as $ M_{max}^{2.8}R_{max}^{-4.8}$, which is rather close to our results.  On the other hand, one can also establish that $\mathcal{E}_{max}$ scales in TPE as $M_{max}^{0.8}R_{max}^{-2.8}$ which is a factor $R_{max}/M_{max}$ different from our results. Similarly, the scalings for Eq. (\ref{eq:power}) are different from the TPE predictions by factors of $(\mathcal{E}_{max}/P_{max})^{1/4}$ and $(P_{max}/\mathcal{E}_{max})^{1/2}$ for $R_{max}$ and $M_{max}$, respectively.   We have no detailed explanation for those results.   If $\hat P_{max}$ is far smaller than 1, which is the Newtonian limit, one instead obtains $R_{max} \propto P_{max}^{1/2} \mathcal{E}_{max}^{-1}$ (a factor differing from our results by $(\mathcal{E}_{max}/P_{max})^{0.2}$,  $M_{max} \propto P_{max}^{3/2} \mathcal{E}_{max}^{-2}$ (differing by $(P_{max}/\mathcal{E}_{max})^{1.3}$) and $P_{max}\propto M_{max}^{2}R_{max}^{-4}$. And as previously discussed, the speed of sound in the TPE approach, which scales as $P_{max}/\mathcal{E}_{max}$, thus effectively scales as $(M_{max}/R_{max})^2$.  In any case, at very high density we expect that $P_{max}\propto\mathcal{E}_{max}$ so that $M_{max}\propto\mathcal{E}_{max}^{-1/2}$ for our results as well as both the Newtonian and TPE cases. This somewhat counterintuitive scaling has been previously noted by \cite{Rhoades1974,Lattimer2011}.
%It should be emphasized that, generally, $M_{max}$ and $R_{max}$ are able to be fit from $\mathcal{E}_{max}$ and $P_{max}$ to relatively better accuracies than the inverse fitting of $\mathcal{E}_{max}$ and $P_{max}$ from $M_{max}$ and $R_{max}$.
 
%These power law fits are able to fit $\mathcal{E}_{max}$ ($P_{max},c_{s,max}$) better (worse) than the fits of Ref. \cite{Cai2023} based on the TPE.   So a combination of these approaches might be judicious.

\section{Fitting and Inverting the Entire $M\mbox{-}R$ Curve\label{sec:inv}}

 Examining the TPE or power-law approaches, it appears there is nothing special about fitting just the maximum mass point.  Furthermore, the accuracy of the TPE approach could improve when applied to configurations less massive than $M_{max}$ since values of the expansion parameters $\hat P$ or $\phi$ at the stars' centers should become smaller.  
 In addition, the constraints $\hat P_c<1/\sqrt{3}$ and $\phi_c<(2\sqrt{e}+4)^{-1}$ at the star's center should be increasingly satisfied for more EOSs as the mass is lowered. 
 In this section, we investigate correlations existing among the quantities $M_{f}, R_{f}, \mathcal{E}_{f}$, $P_{f}$ and $c_{s,f}$ for stars with masses determined by a grid of fractional maximum mass values, $f_i=M_i/M_{max}$, where the energy densities, pressures and sound speeds correspond to the central values of those configurations. 
  If accurate analytic approximations can be found for a variety of values of $f$, we could then determine values for a series of equation of state points ($\mathcal{E}_f$, $P_f$, $c_{s,f}$) thereby effectively achieving an analytical inversion of the $M\mbox{-}R$ curve.   
 The corresponding neutron star equation of state could then be found by interpolation among those equation of state points.  The accuracy of this inversion scheme would depend on the individual accuracies of the correlations at each value of $f$ as well as the number of fractional mass points considered.   

  We begin by arbitrarily choosing a grid $f$ of 11 points on each $M$-$R$ curve corresponding to masses $fM_{max}$, with $f
\in[1,0.95,0.9,0.85,0.8,0.75,2/3,3/5,1/2,2/5,1/3]$.  We then evaluate the corresponding central energy densities $\mathcal{E}_f$, pressures $P_f$ and sound speeds $c_{s,f}$ from solutions of the TOV equation for each EOS in our sample.  

In the TPE approach, we fit, using least squares, correlations resembling those of Eq. (\ref{eq:caifit2}):
\begin{equation}
\nu_f=a_{\nu f}+b_{\nu f} R_f,\qquad\phi_f=a_{\phi f}+b_{\phi f} \left({GM_f\over R_fc^2}\right),\qquad
{c_{s,f}\over c}=\alpha_{cf}+\beta_{cf}\sqrt{GM_f\over R_fc^2}
.\label{eq:caifit3}\end{equation}
Then $\mathcal{E}_{f}=(\phi_f/\nu_f^2){\rm~GeV~fm}^{-3}$
and
\begin{equation}
\hat P_f={1\over3\phi_f^2}\left[\left({1\over2}-2\phi_f\right)-\sqrt{\phi_f^2-2\phi_f+{1\over4}}\right].
\label{eq:pmaxf}\end{equation}

In the power-law approach, we
least-squares fit the data to the functions
 \begin{eqnarray}
     G_{f}&=&a_{Gf}\left({M_{max}\over M_\odot}\right)^{b_{Gf}}\left({R_{f}\over{10\rm~km}}\right)^{c_{Gf}},
%     ,\cr P_{f}&=&a_{P,f}\left({M_{max}\over M_\odot}\right)^{b_{P,f}}\left({R_{f}\over{10~\rm km}}\right)^{c_{P,f}},\cr
%       c_{s,f}&=&a_{c,f}\left({M_{max}\over M_\odot}\right)^{b_{c,f}}\left({R_{f}\over{10~\rm km}}\right)^{c_{c,f}},\cr
%    \hat P_{f}&=&a_{P/\mathcal{E},f}\left({M_{max}\over M_\odot}\right)^{b_{P/\mathcal{E},f}}\left({R_{f}\over{10~\rm km}}\right)^{c_{P/\mathcal{E},f}}.
    \label{eq:powfit1}\end{eqnarray}
where $G\in[\mathcal{E},P,c_s/c,\hat P]$.   Note that we use $M_{max}$ in these fits rather than $M_f=fM_{max}$ since the constant $f$ can be absorbed into the values of the $a_{Gf}$ coefficients.
Parameters and root mean square uncertainties $\delta$ for both the TPE and power-law approaches are given in Table \ref{tab:powfit}.   We note that the power-law approach is more accurate than the TPE approach for predicting $\mathcal{E}$ and $P$ in all cases, and is usually more accurate for predicting $\hat P$ and $c_s$.

Before proceeding, we once again compare the powers of mass and radius appearing in the power-law fits.  Now, examining the averages of the powers for different $f$ values, we see that $\mathcal{E}\propto M^{0.1}/R^{2.5}$, $P\propto M^2/R^4$, and $c_s/c\propto (M/R)^{1.4}$.  Only the fit to the pressure follows dimensional considerations.

The accuracies of fits of these types are fundamentally limited by the fact that a single $(M,R)$ value cannot translate into unique values for $\mathcal{E}_c$ and $P_c$. Those values will depend on the TOV integration through lower masses where the EOSs generally differ.  The fundamental limits to the accuracies, as shown in Table \ref{tab:powfit2}, are in the range of $8-10\%$ and we have determined they cannot be substantially improved by considering higher-order fits.

\begin{figure}
\centering
\includegraphics[width=15cm]{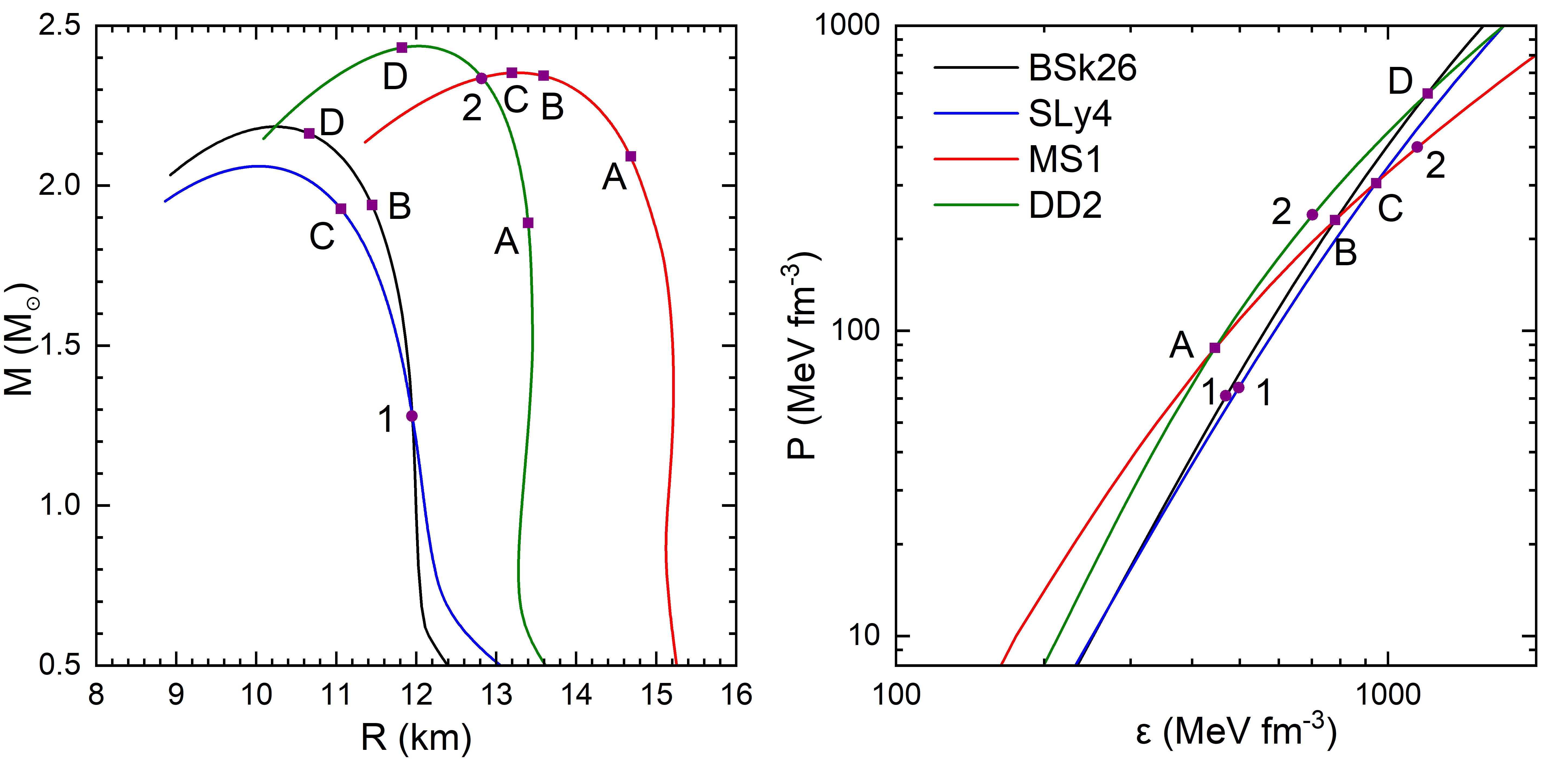}
    \caption{Solid lines show $M\mbox{-}R$ (left panel) and $P\mbox{-}\mathcal{E}$ (right panel) relations for the BSk26 (black), SLy4 (blue), MS1 (red) and DD2 (green) EOSs.  Filled circles with numbers (squares with letters) show where 2 EOSs have the same $M,R$ ($\mathcal{E},P$) pairs.}
\label{fig:cross}\end{figure}

This situation is emphasized in Fig. \ref{fig:cross} which illustrates $M\mbox{-}R$ and $\mathcal{E}\mbox{-}P$ relations for 
the four typical EOSs, BSk26 and SLy4, which are non-relativistic (Skyrme-type), and MS1 and DD2, which are relativistic (RMF-type).   For BSk26 and SLy4, which have similar low-density EOSs, the crossing point (1) in the $M-R$ plane has nearly the same values of $\mathcal{E}_c$ and $P_c$, but for MS1 and DD2, which have rather different low-density EOSs, the crossing point (2) in the $M\mbox{-}R$ plane is characterized by rather different values of $\mathcal{E}_c$ and $P_c$.  At this crossing point, the pressures of DD2 and MS1 are about 25 MeV fm$^{-3}$ and 40 MeV fm$^{-3}$, respectively, and each differs from their geometric mean by about 25\%.   Similarly, the crossing points (A,B,C,D) in the $\mathcal{E},P$ plane are characterized by varying values of $M$ and $R$.   

\begin{table}[h]
 \begin{tabular}{ccc|ccccc} \toprule
    $f=M_f/M_{max}$ & $g_{min}$ & $h_{min}$ & $a_{\mathcal{E}f}$ & $b_{\mathcal{E}f}$ & $c_{\mathcal{E}f}$ & $d_{\mathcal{E}f}$ & $<\delta {\cal E}_f>$  \\ \hline 
     1 & 0.95 &   0.9 &       1.644  &   -0.1408 & -10.46 &    8.614 &    0.00392 \\
0.95 & 0.95 & 4/5 & 0.9003 &      0.1474 &  -5.410 &     3.257 &   0.00251\\
0.9 & 0.95 & 0.9 &0.7642   &   0.2540 &   -3.467 &   1.385 &   0.00220\\
0.85 & 0.95 & 1/2 & 0.6552  &   0.3481 &  -3.249 &     0.9456 &   0.00235\\
4/5 & 0.9 & 1/2 &  0.6094 &  0.4087  &  -3.703  &    1.292 &   0.00224\\
3/4 & 0.85 & 1/2 &  0.5669  &  0.4584 &   -4.113 &  1.606 &   0.00232\\
2/3 & 3/4 & 1/2 & 0.5064 & 0.5279 &-5.259 & 2.731 & 0.00271\\
3/5 & 2/3 & 1/2 & 0.4596 & 0.5818 & -7.108 & 4.525 & 0.00317\\
1/2 & 3/5 & 2/5 & 0.3922 & 0.6497 & -5.948 & 3.303 & 0.00414\\
 2/5 & 1/2 & 2/5 & 0.3282 &  0.7199 &   -9.591 &  6.870 &  0.00761\\
 1/3 & 1/2 & 1/3 & 0.2905 & 0.7503 &  -5.849 &  3.073 &  0.0139\\
 \hline\hline
%\end{tabular}
%\begin{tabular}{ccc|ccccc}
    $f=M_f/M_{max}$ & $g_{min}$ & $h_{min}$  & $a_{Pf}$ & $b_{Pf}$ & $c_{Pf}$ & $d_{Pf}$& $<\delta P_f>$  \\ \hline 
1 & 1 & 3/5 & 0.06712   &  3.103 &      -7.061 &      1.888 &   0.0126\\
0.95 & 0.95 & 3/5 &  0.05614 &   2.639 &   -7.333   &    2.663 &   0.00717\\
0.9 & 0.95 & 2/5 & 0.05091   &  2.517 &    -5.734  &     1.222   &  0.00530\\
0.85 & 0.9 & 2/5 & 0.04693  &  2.408 &    -5.905  &     1.497 &   0.00501\\
4/5 & 0.9 & 1/2 &  0.04181  & 2.337 &     -6.096 &   1.762 &    0.00486\\
3/4 & 4/5 & 2/5 & 0.03708 & 2.271 & -6.328 & 2.056 & 0.00576\\
2/3 & 2/3 & 2/5 & 0.03109    & 2.145 &    -7.776 &  3.636 &    0.00547\\
3/5 & 3/5 & 2/5 & 0.02506 & 2.108 & -9.014 & 4.912 & 0.0251\\
1/2 & 1/2 & 2/5 & 0.01768 & 2.057 & -14.05 & 9.998 & 0.0110\\
 2/5 & 1/2 & 1/3 &  0.01120 &       2.000 & -8.081 & 4.073 & 0.0216\\
 1/3 & 1/2 & 1/3 & 0.009059 &       1.944 &  -6.854 &     2.899 & 0.0313\\ 
\hline\hline
    $f=M_f/M_{max}$ & $g_{min}$ & $h_{min}$  & $a_{cf}$ & $b_{cf}$ & $c_{cf}$ & $d_{cf}$& $<\delta c_{s,f}>$  \\ \hline 
1 & 1 & 2/5 & 0.1963  &  2.025 &      -3.093 &   0.9257 &   0.0574\\
0.95 & 0.95 & 2/3 &  0.2397 &  1.810  &  -3.480   &  1.640 &   0.00721\\
0.9 & 4/5 & 2/3 & 0.2808   &  1.550 &   -6.027  &     4.373   &  0.00554\\
0.85 & 2/3 & 1/2 & 0.2992  & 1.544   &   -10.38  &   8.856 &   0.00539\\
4/5 & 3/5 & 1/2 & 0.3027  & 1.469   &    -6.363 &   4.944 &    0.00634\\
3/4 & 1/2 & 2/5 & 0.3008 &       1.403 &  -5.594 &     4.265 & 0.00768\\
2/3 & 1/2 & 1/3 & 0.2827  & 1.392 &      -3.090 &  1.813 &    0.0116\\
3/5 & 2/5 & 1/3 & 0.2746 &1.300 &-5.595 & 4.362 & 0.0155\\
1/2 & 0.95 & 0.9 & 0.3892 & 0.9500 & 10.69 & -11.77 & 0.0180\\
2/5 & 0.9 & 0.85 & 0.3791 &      0.7748 &  14.78 &  -15.73 & 0.0210\\
1/3 & 0.85 & 3/4 & 0.3558 &      0.6507 & 8.830 & -9.694 & 0.0235\\
\hline\hline
    $f=M_f/M_{max}$ & $g_{min}$ & $h_{min}$  & $a_{\hat Pf}$ & $b_{\hat Pf}$ & $c_{\hat Pf}$ & $d_{\hat Pf}$& $<\delta \hat P_f>$  \\ \hline 
1 & 1 & 1/2 & 0.04385 & 3.234 &-4.366 & 1.048 & 0.0136\\
0.95 & 0.95 & 2/5 & 0.06538 & 2.504 & -3.233 & 0.7212 & 0.00681\\
0.9 & 0.85 & 2/5 & 0.07165 & 2.231 & -3.223 & 0.9878 & 0.00519\\
0.85 & 3/4 & 2/5 & 0.07192 & 2.047 & -3.379 & 1.332 & 0.00452\\
4/5 & 3/5 & 1/2 & 0.06947 & 1.914 & -7.931 & 6.021 & 0.00414\\
 3/4 & 3/5 & 2/5 &  0.06506 &      1.810 & -4.055 &  2.256 & 0.00441\\
 2/3 & 1/2 & 2/5 & 0.05577 & 1.670 & -5.921 & 4.271 & 0.00546\\
3/5 & 1/2 & 1/3 & 0.05176 & 1.578 &  -3.422 &1.860 & 0.00785\\
1/2 & 2/5 & 1/3 & 0.04410 & 1.426 & -5.495 & 4.063 & 0.0120\\ 
2/5 & 0.9 & 4/5 & 0.04351 &       1.164 & 4.515 &  -5.778 &
     0.0135\\
1/3 & 4/5 & 3/4 &  0.03624 &       1.059 & 10.36 & -11.45 &
     0.0146\\
     \hline\end{tabular}

\caption{Parameters and RMS errors $<\!\!\delta G\!\!>$ for the fits in Eq. (\ref{eq:eqfit1}). $a_{\mathcal{E}f}$ and $a_{Pf}$ have units of GeV fm$^{-3}$.  Equation of state models considered were restricted to those satisfying $M_{max}\ge2M_\odot$.  $g_{min}$ and $h_{min}$ refer to the two fractional $M_{max}$ radii that give the minimum RMS error for the indicated fractional $M_{max}$ point $f$.  \label{tab:powfit1}}
\end{table}

\begin{figure}
\includegraphics[width=9cm]{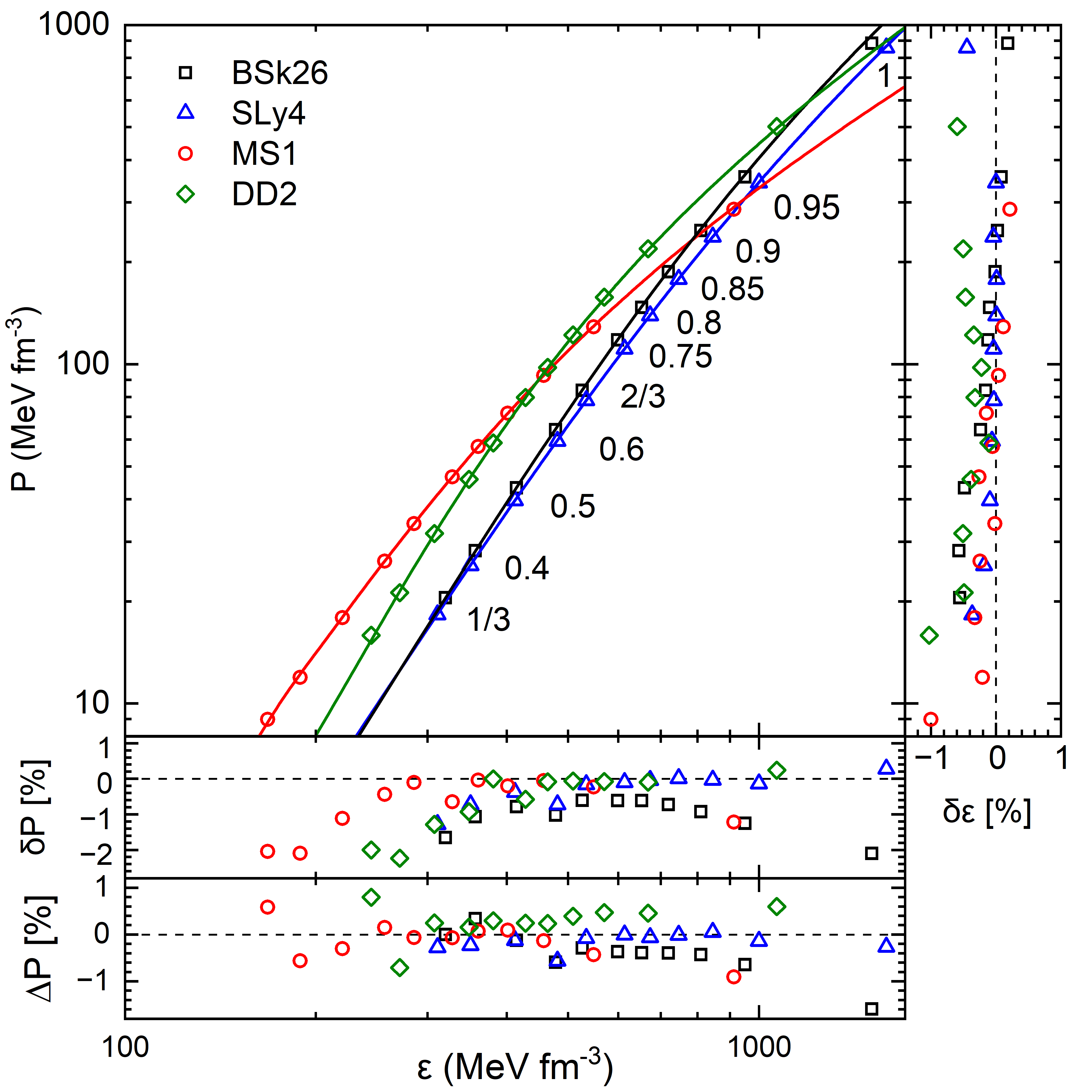}
\includegraphics[width=9cm]{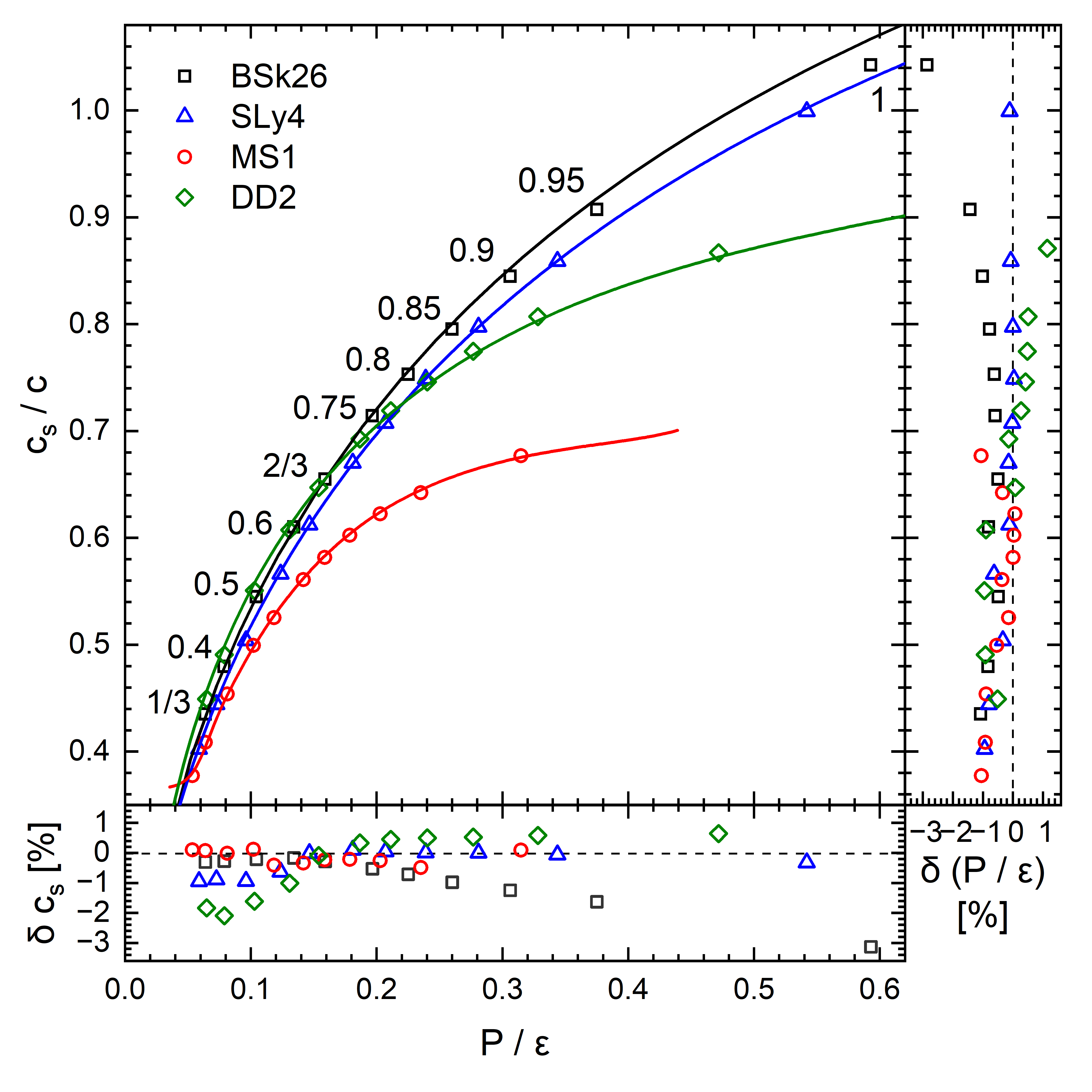}
    \caption{Fidelity of two-radius power-law fits fractional maximum mass inversion technique. Left panel:  Solid lines show the $P\mbox{-}\mathcal{E}$ relations for  BSk26, SLy4, MS1 and DD2 EOSs.  Points show $\mathcal{E}_c$ and $P_c$ values reconstructed from Eq. (\ref{eq:G2}) at the indicated  $M_{max}$ fractions.  The side and lower panels show logarithmic errors at each $M_{max}$ fraction.  The lowest panel shows the true deviation $\Delta P$ from the EOS, defined in Eq. (\ref{eq:DeltaP}). Right panel: Solid lines show the $c_s/c\mbox{-}P/\mathcal{E}$ relation, while points show the reconstructed values of $c_{s,c}$$/c$ and $P_c/\mathcal{E}_c$ from Eq. (\ref{eq:G2}). The side and lower panels show logarithmic errors.}
\label{fig:2Rcomp}\end{figure}

Including additional $M\mbox{-}R$ information can break this degeneracy and render both approaches more accurate.  \cite{Ofengeim2023} suggested including the radius $R_{f=1/2}$ at half the maximum mass in addition to that at the maximum mass $R_{f=1}$ as a fit variable.  Here, we go further, and show that substantially greater accuracies can be achieved by optimizing the choices of the two radii.  For the remainder of this paper, we focus on the power-law approach instead of the TPE approach due to its inherently greater accuracy.   

Using the available grid $f$ of masses and radii, we find that fitting for the EOS for a given mass grid point $f$ using an optimized selection of two radius grid values ($g\in f,h\in f$) increases the RMF accuracies for both $\mathcal{E}$ and $P$ to better than 1\% and often better than 0.5\% (see Table \ref{tab:powfit1}).  $g$ and $h$ may or may not include $f$.
To be specific, we determined fits of the quantities
\begin{eqnarray}
    G_{f}&=&a_{Gf}\left({M_{max}\over M_\odot}\right)^{b_{Gf}}\left({R_{g}\over10{\rm~km}}\right)^{c_{Gf}}\left({R_{h}\over10{\rm~km}}\right)^{d_{Gf}},
  \label{eq:eqfit1}\end{eqnarray}
  where the $a$'s, $b$'s, $c$'s and $d$'s  are fitting parameters, by minimizing the quantities
  \begin{equation}
\chi_{Gf}^2=\sum_i^N\left[\ln\left({G_{fi}\over a_{Gf}}\right)-b_{Gf}\ln\left({M_{max,i}\over M_\odot}\right)-c_{Gf}\ln\left({R_{gi}\over{10\rm~km}}\right)-d_{Gf}\ln\left({R_{hi}\over10{\rm~km}}\right)\right]^2
 \label{eq:G2}     
  \end{equation}
with respect to the parameters $[g,h,a,b,c,d]$.   We used the shorthand that, although the coefficients $[a,b,c,d]$ depend on $f,g$ and $h$, the optimized parameters are referred to simply as [$a_{Gf},b_{Gf},c_{Gf},d_{Gf}$], and are given in Table \ref{tab:powfit1}.  
  
It is clear that the accuracy of this method is far superior to the previous power-law fits based on one radius.  The average accuracy for $\mathcal{E}$ is better than 0.4\% except for the cases $1/3\le f\le1/2$, and that of $P$ is better than 0.6\% except for $1/3\le f\le3/5$, which both represent an order-of-magnitude increase in precision.
Again comparing the powers of mass and radius appearing in these fits, and summing the two radius exponents, we see that $\mathcal{E}\propto M^{0.5}/R^{2.5}$ and $P\propto M^2/R^4$.

To illustrate the potential of this method, we selected four typical, but dissimilar, EOSs, the same ones used in Fig. \ref{fig:cross}.  Fig. \ref{fig:2Rcomp} compares their $\mathcal{E}{\mbox-}P$ relations with reconstructions from their corresponding $M{\mbox-}R$ curves using Eq. (\ref{eq:eqfit1}) with the parameters tabulated in Table \ref{tab:powfit1}.  Deviations are generally less than 1\%, with the largest deviations generally connected with the lower mass points.  Similar inversion accuracies are found for $\hat P=P/\mathcal{E}$ and $c_s/c$, as is shown in Fig. \ref{fig:2Rcomp}.
  
  Actually, from the point of view of reproducing the actual EOS curves, the fits are even more accurate than these deviations, $\delta\mathcal{E}, \delta P, \delta\hat P$ and $\delta c_s$ would suggest.  Note from Fig. \ref{fig:2Rcomp} that the reconstructed $P_{fit,i}$ at a point $f_i$ often lie close to the actual EOS curves even though the $\delta P$ values appear to be relatively large.  To measure the true deviation from the EOS curve, we define the quantity $\Delta P$
  \begin{equation}
      \Delta P_i = {P_{fit,i}(\mathcal{E}_{fit,i})\over P_i(\mathcal{E}_{fit,i})}-1, \label{eq:DeltaP}
  \end{equation}
  where $P_i$ is the actual pressure at the energy density $\mathcal{E}_{fit,i}$ that is reconstructed from the fit at $M_{f_i}$, and $P_{fit,i}$ is the fitted pressure at $M_{f_i}$.  This quantity is shown in the lowest subpanel of the left panel of Fig. \ref{fig:2Rcomp}. 

\begin{table}[h]
 \begin{tabular}{ccc|cccccc} \toprule 
   $f=M_f/M_{max}$ & $g_{min}$ & $h_{min}$ & $a_{nf}$ & $b_{nf}$ & $c_{nf}$ & $d_{nf}$ & $<\delta n_f>$  \\ \hline 
   1 &   0.95 &  0.9 & 1.700  &   -0.5096 & -6.894 &  5.434 &    0.00874 \\
0.95 & 0.95 & 0.9 & 0.8759 &  -0.00952 &  -8.223 &  6.248 &   0.00863\\
0.9 & 0.95 & 3/4 &0.7409   &  0.1457 &   -3.809 &   1.671 &   0.00821\\
0.85 & 0.95 & 1/2 & 0.6619  &   0.2498 &  -2.910 &  0.6677 &   0.0106\\
4/5 & 0.95 & 1/3 &  0.5844 &  0.3465  &  -2.700  &  0.3725 &   0.0111\\
3/4 & 0.9 & 2/5 &  0.5532  &  0.4045 &   -3.033 &  0.6411 &   0.0103\\
2/3 & 0.85 & 1/3 & 0.4877 & 0.5016 &-3.104 & 0.6215 & 0.0104\\
3/5 & 3/4 & 2/5 & 0.4565 & 0.5506 & -3.847 & 1.307 & 0.00908\\
1/2 & 3/5 & 2/5 & 0.4041 & 0.6099 & -5.543 & 2.939 & 0.00816\\
2/5 & 1/2 & 2/5 & 0.3405 &  0.6861 &   -8.978 & 6.294 &  0.00939\\
1/3 & 1/2 & 1/3 & 0.3025 &   0.7220 &  -5.553 & 2.810 &    0.0142\\
 \hline\hline 
%\end{tabular}
%\begin{tabular}{ccc|cccccc}
    $f=M_f/M_{max}$ & $g_{min}$ & $h_{min}$  & $a_{\mu f}$ & $b_{\mu f}$ & $c_{\mu f}$ & $d_{\mu f}$& $<\delta \mu_f>$ 
    & $<\delta n_f>$ 
    \\ \hline 
1 & 1 & 2/5 & 0.7308   &  1.308 &   -1.851 &      0.4070 &   0.00990 & 0.00873
\\
0.95 & 0.95 & 2/5 &  0.9405 &  0.7278 &  -1.154   & 0.4070 &   0.00409 & 0.00862
\\
0.9 & 0.95 & 1/3 & 0.9590   & 0.5793 & -0.8390  &   0.2521  &  0.00315 & 0.00820
\\
0.85 & 4/5 & 1/2 & 1.001  & 0.4669 &  -1.167  &  0.6878 &   0.00255 & 0.0106
\\
4/5 & 3/4 & 1/2 & 1.056  & 0.3974 &  -1.137 &  0.7309 &    0.00234 & 0.0111
\\
3/4 & 3/4 & 2/5 & 1.007 & 0.3410 & -0.7358 & 0.3881 & 0.00346 & 0.0103
\\
2/3 & 3/5 & 1/2 & 1.006    & 0.2685 &  -1.632 &  1.358 &    0.00263 & 0.0104
\\
3/5 & 3/5 & 2/5 & 1.001 & 0.2244 & -0.711 & 0.4820 & 0.00313 & 0.00906
\\
1/2 & 1/2 & 2/5 & 0.9930 & 0.1712 & -0.9311 & 0.7551 & 0.00330 & 0.00814
\\
2/5 & 1/2 & 1/3 & 0.9845 & 0.1270 & -0.4195 & 0.2870 & 0.00259 & 0.00937
\\
1/3 & 1/2 & 1/3 & 0.9787 & 0.1022 & -0.3245 & 0.2170 & 0.00233 & 0.0142
\\ 
     \hline\end{tabular}

\caption{The same as Table \ref{tab:powfit1} but for the baryon number density $n$ and chemical potential $\mu$.  $a_{nf}$ and $a_{\mu f}$ have units of fm$^{-3}$ and GeV, respectively. The last column in the $\mu$ section shows the RMS error $<\!\!\delta n_f\!\!>$ where $n_f=(\mathcal{E}_f+P_f)/\mu_f$.
   \label{tab:powfit2}}
\end{table}

\hspace*{-1.cm}
\begin{figure}[h]
    \includegraphics[width=9cm]{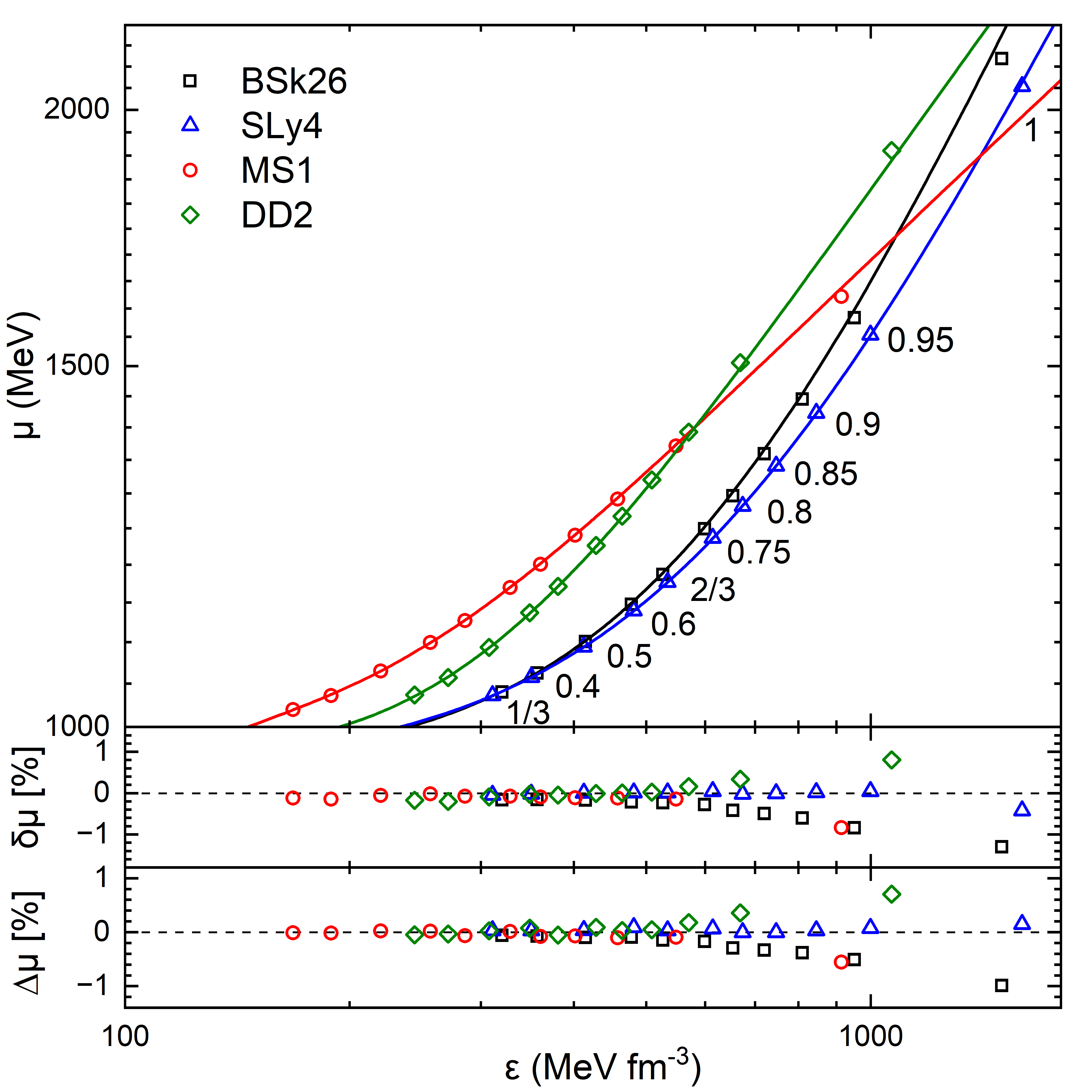}
    \includegraphics[width=9cm]{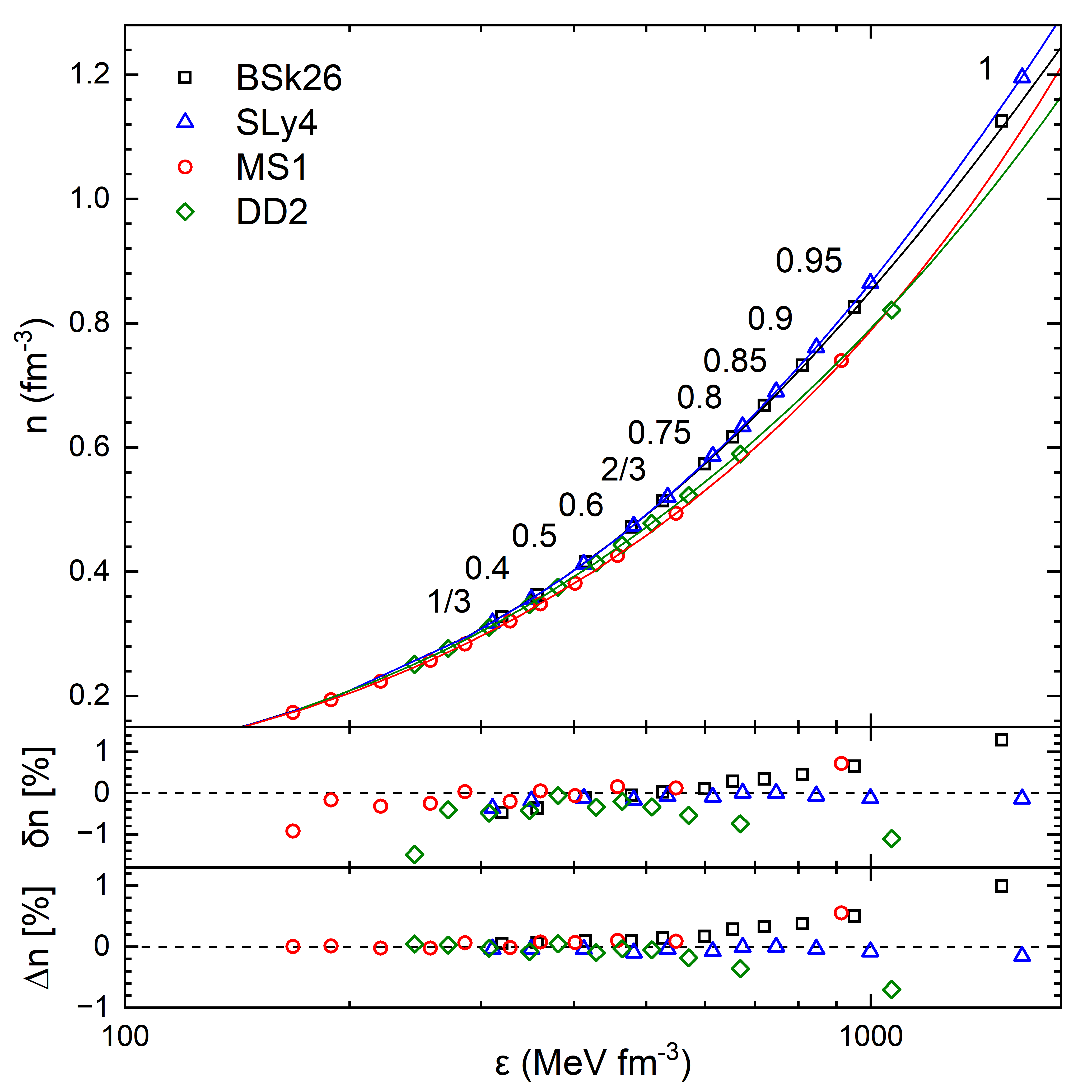}
    \caption{The same as  Fig. \ref{fig:2Rcomp} except for density $n$ and chemical potential $\mu$ as functions of energy density $\mathcal{E}$.}
\label{fig:2Rcomp2}\end{figure}

In some contexts, one may wish to infer the central baryon chemical potential $\mu$ and/or the baryon number density $n$ directly from the $M{\mbox-}R$ curve.  We therefore computed fit parameters for $\mu_f$ and $n_f$ using Eq (\ref{eq:eqfit1}),
%, but the fit to $n$ using the identity $n=(\mathcal{E}+P)/\mu$ with the fitted values for $\mathcal{E},P$ and $\mu$ has a smaller RMS error $<\delta n_f>$ for all $f$ than the direct fit to $n$ using Eq. (\ref{eq:eqfit1}).  We provide fitting coefficients for $\mu$ 
and also show them and the resulting RMS errors in Table \ref{tab:powfit2}.    It is interesting that the RMS deviations in $\mu$ are less than 0.5\% except at low masses, and those in $n$ are generally even smaller.  Both are generally smaller than those of $\mathcal{E}$ or $P$ or even $P/\mathcal{E}$. Fig. \ref{fig:2Rcomp2} compares $\mu$ and $n$ for the four standard equations of state used in Fig. \ref{fig:cross} with their corresponding $M\mbox{-}R$ inversions obtained from Eq. (\ref{eq:eqfit1}) using the coefficients tabulated in Table \ref{tab:powfit2}.

\begin{figure}
    \centering
    \includegraphics[width=0.7\linewidth]{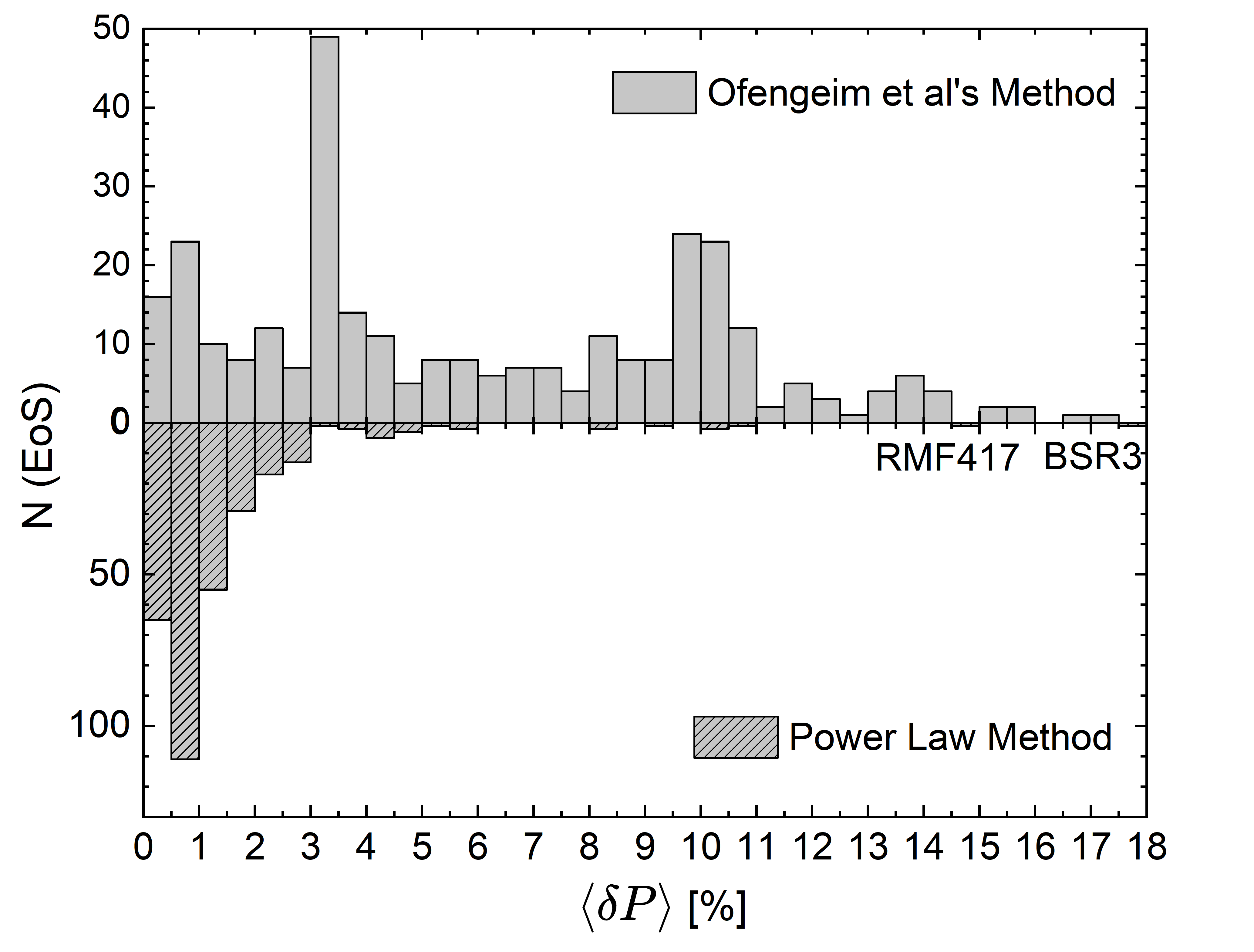}
    \caption{Histogram showing the number of EOSs with particular ranges of pressure RMS errors at the 11 fiducial points $f$ along the $M\mbox{-}R$ curves for the 312 EOSs in the tabulation of \cite{Sun2024} for which $M_{max}\ge2M_\odot$. Upper portion shows results for the method of \cite{Ofengeim2023}; lower portion shows results for the power-law method of this paper, identifying specific EOSs in that tabulation having large errors. }
    \label{fig:hist}
\end{figure}
Although it was not the primary goal of Ref. \citep{Ofengeim2023} to predict central pressures and energy densities of stars along the $M-R$ curve, their methods allow one to do so.  It is useful to compare the results so obtained with ours.  By inverting their correlations
\begin{equation}
    G_i=c_i
    \left({P_{max}\over\mathcal{E}_{max}}\right)^{p_i}
    \left({\mathcal{E}_{max}\over\mathcal{E}_0}\right)^{q_i}
    +d_i,
\label{eq:Ofengeim}\end{equation}
where $G_i\in[M_{max},R_{max},c^2_{s,max}/c^2]$,  $(c_i,d_i,p_i,q_i)$ are fitting parameters, and $\mathcal{E}_0\simeq0.150$ GeV fm$^{-3}$ is the nuclear saturation energy density, one is able to determine $P_{max},\mathcal{E}_{max}$ and $c^2_{s,max}/c^2$ from $M_{max}$ and $R_{max}$ for a given EOS.  Combining this with a semi-universal approximation they proposed, valid for $\mathcal{E}>3\mathcal{E}_0$, of the form
$P(\mathcal{E},P_{max},\mathcal{E}_{max},c_{s,max})$ [Eqs. (2a,2b) in \cite{Ofengeim2023}], a relation of the form $P(\mathcal{E},P_{max},\mathcal{E}_{max})$ can be obtained.  Fig. \ref{fig:hist} displays a histogram comparing the RMS accuracies of this approach at our 11 mass points $f$ to those from our approach using Eq. (\ref{eq:eqfit1}) for the set of 316 EOSs in the tabulation of \cite{Sun2024} for which $M_{max}\ge2M_\odot$.  About 2/3 of these interactions have RMS pressure errors less than 1\% with our approach, compared with about 1/7 using the approach of \cite{Ofengeim2023}.  A code has been placed in the Zenodo repository \citep{zenodo_TOV} to invert complete, arbitrary $M{\mbox-}R$ curves to give their underlying $\mathcal{E}\mbox{-}P$ relations.

Since the sample \citep{Sun2024}  of EOSs used to establish our fits does not include any with first-order phase transitions, it can be expected that our method may not give good results for those particular EOSs.  This is elaborated in \S \ref{sec:discuss}.  However, as we also discuss in \S \ref{sec:discuss}, the accuracy of our method can arbitrarily improved, even in the case of first-order phase transitions, by using the initial guesses it provides together with Newton-Raphson iterations with TOV integrations utilizing $(M_f,R_f)$ information but not any EOS information at densities above the star's core-crust transition.
%It should be emphasized that although a particular $M-R$ curve corresponds to a unique $\mathcal{E}-P$ relation, the maximum mass point $M_{\rm max}-R_{\rm max}$ does not denote unique values for $\mathcal{E}_{\rm max}$ and $P_{\rm max}$, but rather approximate values with uncertainties of order a few percent.  Those estimates can be made more precise, however, by factors of several, if additional fractional $M_{\rm max}$ points are considered. 

\section{From $M\mbox{-}R$ to $P\mbox{-}\mathcal{E}$ \label{sec: ptop}}

The analytic inversion of the $M\mbox{-}R$ curve discussed in \S \ref{sec:inv} cannot be applied unambiguously to observational $M-R$ data because it requires knowledge of, or a probability distribution for, $M_{max}$.  This extra uncertainty could be removed, however, if accurate analytic correlations can be found that relate radii at fixed mass values along $M\mbox{-}R$ curves to their corresponding central energy densities and pressures.  

A first step toward this goal is to develop fits for a fixed grid of $M$ values, which we take as $M_j\in[2.0,1.9,1.8,1.7,1.6,1.5,1.4,1.3]M_\odot$; as before, we will only consider equations of state that satisfy $M_{max}\ge2M_\odot$.  

First we develop correlations for single fixed mass values, using data from the 316 equations of state from the tabulation of Ref. \citep{Sun2024}.  Specifically, for the TPE approach, we assume
\begin{equation}
\nu_j=a_{\nu}+b_{\nu} R_j,\qquad\phi_j=a_{\phi}+b_{\phi} \left({GM_j\over R_jc^2}\right),\qquad
\frac{c_{s}}{c}=\alpha_{c}+\beta_{c}\sqrt{GM_j\over R_jc^2}
.\label{eq:caifit4}\end{equation}
Then
\begin{equation}
\frac{\mathcal{E}_j}{{\rm GeV~fm}^{-3}}={\phi_j\over\nu_j^2},\qquad \hat P_j={1\over3\phi_j}\left[\left({1\over2}-2\phi_j\right)-\sqrt{\phi_j^2-2\phi_j+{1\over4}}\right].
\label{eq:pmaxj}\end{equation}
In the power-law approach, we
least-squares fit the data to the functions
 \begin{eqnarray}
     G_{j}&=&a_{G}\left({R_{j}\over{10\rm~km}}\right)^{b_{G}},
    \label{eq:powfit3}\end{eqnarray}
Note that the fit does not include a mass term since that only introduces a constant correction to the $a_{G}$ parameters.  Results are given in Table \ref{tab:powfit3}.

\begin{table}
      \begin{tabular}{c|ccc|cccc} \toprule
  $M_j/M_\odot$ & $a_{\mathcal{E}}$ & $b_{\mathcal{E}}$ 
    & $<\delta\mathcal{E}_j>$ & $a_{\nu}$ & $b_{\nu}$ & $<\delta\mathcal{E}_{j,TPE}>$ & $<\delta\nu_{j,TPE}>$
    \\ \hline
  2.0  & 1.487 & -3.557 &  0.0645
  & -0.1751 & 0.04712 & 0.0956  & 0.0424
  \\
 1.9& 1.343  & -3.429 & 0.0597 
 & -0.1409 & 0.04490 & 0.08768&0.0392
 \\
 1.8  & 1.235  & -3.429 & 0.0568
 &-0.1166 & 0.04333 & 0.0829 &0.0380
 \\
 1.7  & 1.142  & -3.288 & 0.0546
 & -0.09710 & 0.0205 & 0.0798 &0.0368
 \\ 
1.6 & 1.0600   &  -3.324   & 0.0533 
& -0.08050 & 0.04093 & 0.0778 &0.0362
\\
1.5 & 0.9850  & -3.195   & 0.0520
& -0.05371 & 0.03903 & 0.0759 &0.0352
\\
1.4 & 0.9157 & -3.158   & 0.0520
& -0.05371 & 0.03903 & 0.0759 &0.0352
\\ 
1.3 & 0.08501 & -3.124  & 0.0520
& -0.04281 & 0.03819 & 0.07587 &0.0350
\\
\hline\hline
%\end{tabular}    
%       \begin{tabular}{c|ccc|ccc} \toprule
    $M_j/M_\odot$ & $a_{P}$ & $b_{P}$  & $<\delta P_j>$ 
   & $a_{\phi}$ & $b_{\phi}$ & $<\delta P_{j,TPE}>$ &$<\delta\phi_{j,TPE}>$
    \\ \hline 
  2.0  & 0.6404 & -5.773 & 0.0940 & 0.02092 & 0.3957 & 0.1079 & 0.00711
  \\
 1.9  & 0.4914  & -5.541  & 0.0877
 & 0.01428 & 0.4277 & 0.0930 &0.00737
 \\
 1.8& 0.3923 & -5.321  & 0.0845 & 0.01042 & 0.4479 & 0.0869 &0.00791
 \\
 1.7  & 0.3182 & -5.051  & 0.0824
 & 0.007803 & 0.4631 & 0.0833& 0.00860
 \\ 
1.6 & 0.2604 & -4.909  &0.0817 & 0.004764 & 0.4844 &0.0818 & 0.00930
\\
1.5 & 0.2140  & -4.786  & 0.0822 
& 0.003789 & 0.4844 & 0.0813 &0.0102
\\
1.4 & 0.1762  & -4.681  & 0.0822 
& 0.003989 & 0.4930 & 0.0817  &0.0112
\\ 
1.3 & 0.1447 & -4.579  & 0.0835 & 0.003334 & 0.4988 & 0.0830 &0.0122  
\\
\hline\hline
%\end{tabular}    
%
%      \begin{tabular}{c|cccc|ccc} \toprule
   $M_j/M_\odot$ & $a_{c}$ & $b_{c}$ & $<\delta c_{s}>$
  & $\alpha_{c}$ & $\beta_{c}$& $<\delta c_{s,j,TPE}>$
    \\ \hline 
  2.0  & 0.8879  & -0.7471 & 0.0716 & 0.1457 &2.576& 0.0712
  \\
 1.9  & 0.8414 & -0.6894 & 0.0656 &0.1857 &2.399 & 0.0650 
 \\
 1.8& 0.8018  &-0.6530  & 0.0623 &0.2067 &2.300 & 0.0620
 \\
 1.7  & 0.7661  & -0.6287  & 0.0603&0.2173&2.245 & 0.0597
 \\ 
1.6 & 0.7331 & -0.6129  & 0.0579 &0.2211 &2.224 & 0.0583
\\
1.5 & 0.7022 & -0.6043 & 0.0577  &0.2198 &2.233 & 0.0573
\\
1.4 & 0.6173  & -0.6036 & 0.0569  &0.2131 & 2.279& 0.0566 
\\ 
1.3 & 0.6446   & -0.6057 & 0.0563 & 0.2047& 2.343& 0.0560 
\\
\hline\hline
 $M_j/M_\odot$ & $a_{\hat P}$ & $b_{\hat P}$ & $<\delta\hat P_j>$
  &  & & $<\delta\hat P_{j,TPE}>$
    \\ \hline 
2.0 &0.4038 &-2.216& 0.0358 & && 0.0262\\
1.9&0.3656 &-2.012 & 0.0284 && & 0.0177\\
1.8&0.3177 & -1.872 & 0.0253 & && 0.0175\\
1.7& 0.2786&-1.763 &0.0236& & &0.0181\\
1.6& 0.2457&-1.671& 0.0225 & & &0.0186\\
1.5&0.2172&-1.592&0.0218 & & &0.0190\\
1.4& 0.1924 & -1.523  & 0.0217 && & 0.0197\\
1.3 & 0.1701 & -1.456 & 0.0217 & & &0.0203\\
\hline\hline
\end{tabular}    

\caption{The parameters and RMS uncertainties $<\!\!\delta G\!>$ for the power-law fits in Eq. (\ref{eq:powfit3}) are in the center blocks and those for the TPE fits in Eqs. (\ref{eq:caifit4}) and (\ref{eq:pmaxj}) are in the right blocks.
$a_{\mathcal{E}}$ and $a_{P}$ have units of GeV fm$^{-3}$ and $b_{\nu}$ has units of km$^{-1}$.}
     \label{tab:powfit3}
\end{table}

The results have only moderate precision, and the power-law fits appear to be generally better than those from the TPE approach.  However, it is interesting that, in the TPE approach, the auxiliary functions $\phi$ and $\nu$ are fit with better precision than $\mathcal{E}$ or $P$.  As we will see, this is connected to the strong correlation between the predicted values of $\mathcal{E}$ and $P$, such that $\hat P=P/\mathcal{E}$ is also fit to higher accuracy than $P$ or $\mathcal{E}$ alone. This variable could be used together with $\mathcal{E}$
%another, such as $\omega= P/\mathcal{E}^3$
, to infer $P$ and $\mathcal{E}$.
%\begin{equation}
%    \mathcal{E}={\hat P\over\omega^2},\qquad P=\left({\hat P\over\omega}\right)^2.
 %   \end{equation}
In fact, the variable pair $\mathcal{E}\mbox{-}\hat P$ has the advantage that it does not have a real-value constraint as does the $\phi\mbox{-}\nu$ pair.   In practice, however, fitting $\mathcal{E}_c$ and $\hat P_c$ produces the smallest net uncertainties in $\mathcal{E}_c$ and $P_c$ of any other combination.

Unfortunately, these fitting formulae have limited applicability due to their large mass grid spacing.  This can be somewhat alleviated by obtaining general fits that transform an arbitrary $M-R$ point into its corresponding $\mathcal{E}_c$ and $P_c$ (or $\hat P_c$, etc.).  This is possible since the coefficients $a$ and $b$ in Table \ref{tab:powfit3} vary more or less smoothly along the mass grid.  
This smooth behavior for all $G\in[\mathcal{E}/({\rm GeV~fm}^{-3}),P/({\rm GeV~fm}^{-3}),P/\mathcal{E},c_s/c,\phi,\nu]$ variables suggests a general fitting formula of the form
\begin{equation}
    \ln G=a_G+b_G\ln\left(\frac{M}{M_\odot}\right)+c_G\ln\left(\frac{R}{{\rm km}}\right)+d_G\left[\ln \left(\frac{M}{M_\odot}\right)\right]^2+e_G\ln \left(\frac{M}{M_\odot}\right)\ln \left(\frac{R}{{\rm km}}\right)+f_G\left[\ln\left(\frac{R}{{\rm km}}\right)\right]^2.
\label{eq:powfit4}\end{equation}
  Using the 316 equations of state from the tabulation of \cite{Sun2024} at the 8 fixed mass points previously utilized, least squares best fitting parameters for Eq. (\ref{eq:powfit4}) were obtained and are given in Table \ref{tab:genfit}.  It is again noticeable that fits of $\hat P,\nu$ and $\phi$ are much more accurate than those of $\mathcal{E}$ or $P$.

Nevertheless, as we have already discussed, such fitting formulae are fundamentally limited by the fact that an $M,R$ point does not translate into unique values for $\mathcal{E},P$.  To improve the accuracy, more information from the $M\mbox{-}R$ relation has to be incorporated into these fits.  One possibility is to utilize the inverse slope $dR/dM$ at an $(M,R)$ point, which could be incorporated as follows:
\begin{eqnarray}
   \ln G=a_G&+&b_G\ln\left(\frac{M}{M_\odot}\right)+c_G\ln\left(\frac{R}{{\rm km}}\right)+d_G\left[\ln \left(\frac{M}{M_\odot}\right)\right]^2+e_G\ln \left(\frac{M}{M_\odot}\right)\ln \left(\frac{R}{{\rm km}}\right)\nonumber\\
    &+&f_G\left[\ln\left(\frac{R}{{\rm km}}\right)\right]^2+g_G \left(\frac{dR}{dM}\right)\left(\frac{M_\odot}{{\rm~km}}\right).
\label{eq:powfit5}\end{eqnarray}
We repeated the fitting procedure using this modified formula with the results shown in Table \ref{tab:genfit1}. This general inversion moderately improves the accuracies all quantities.   It is again noticeable that fits of $\hat P,\nu$ and $\phi$ are much more accurate than those of $\mathcal{E}$ or $P$.

\begin{table}
      \begin{tabular}{c|cccccc|c} \toprule
      $G$ & $a_G$ & $b_G$ & $c_G$ & $d_G$ & $e_G$ & $f_G$ & $<\delta G>$\\\hline
$\mathcal{E}$&  2.304 & 2.889 & -0.04210 & 0.5760 & -0.9289 &     -0.5079 &    0.0664\\
      $P$& 14.36 &      7.496 &     -10.19 &      1.317 &     -2.358 &      1.242 &    0.0721\\
$\hat P$& 2.014 &      4.608 &     -10.15 &     0.7409 &     -1.429 &      1.750 &    0.0186\\
%$\omega$&  4.948 &      1.037 &     -5.184 &    0.09736 &    -0.33251 &      1.172 &    0.0329\\
      $c_s/c$&   15.14 &     0.08732 &     -11.39 &    0.07527 &     0.1732 &      2.025 &    0.0458\\
      $\phi$&   0.6581 &    -0.1625 &     -1.581 &    -0.09261 &     0.4288 &     0.08617 &   0.00948\\
      $\nu$&   -0.8094 &     -1.526 &    -0.7696 &    -0.3343 &     0.6789 &     0.2970 &    0.0328\\
      \hline\hline
      \end{tabular}

\caption{The parameters and root mean square uncertainties $<\!\!\delta G\!\!>$ for Eq. (\ref{eq:powfit4}). All parameters are dimensionless. The units of $\mathcal{E}$ and $P$ are in GeV fm$^{-3}$.}
 \label{tab:genfit}
 \end{table}

\begin{table}
      \begin{tabular}{c|ccccccc|c} \toprule
      $G$ & $a_G$ & $b_G$ & $c_G$ & $d_G$ & $e_G$ & $f_G$ & $g_G$ &$<\delta G>$\\\hline\\[-10pt]
$\mathcal{E}$&   0.8001 &     -1.485 &      1.574 &    -0.4902 &      1.012 &    -0.9244
&    -0.07980 &    0.0360\\ 
 $P$ &   12.76 &      2.834 &     -8.469 &     0.1803 &    -0.2884 &     0.7983
&    -0.08507 &    0.0406\\
$\hat P$ &  11.96 &     4.319 &    -10.04 &     0.6705 &     -1.301 &      1.723 &  -0.005267
&     0.0186\\
%$\omega$ &    6.053 &      3.080 &     -6.246 &     0.6061 &     -1.241 &      1.424&     0.03852 &    0.0329\\
$c_s/c$&    15.87 &      2.190 &     -12.17 &     0.5878 &    -0.7600 &      2.225
&     0.03836 &    0.0458\\
$\phi$ &   0.6862 &    -0.1593 &     -1.582 &    -0.09183 &     0.4274 &     0.08648
&   5.832e-5   & 0.00995\\
$\nu$ &   -0.05693 &     0.6629 &     -1.578 &     0.1992 &    -0.2925 &     0.5054
&     0.03993 &    0.0328\\
      \hline\hline
      \end{tabular}
      \caption{The same as Table \ref{tab:genfit} except for Eq. (\ref{eq:powfit5}).}
 
 \label{tab:genfit1}
 \end{table}
%{\color{red}We need another figure to show the performance of individual equations of state when including $dR/dM$, similar to Fig. 1.}

For the four EOSs utilized in the comparison in Fig. \ref{fig:cross}, we repeat the comparisons of inversions for $P_c\mbox{-}\mathcal{E}_c$ and $c_{s,c}/c$$\mbox{-}\hat P$ for the fits of Eq. (\ref{eq:powfit4}) in Fig. \ref{fig:2Rcomp1} for the reference masses.
\begin{figure}
    \includegraphics[width=9cm]{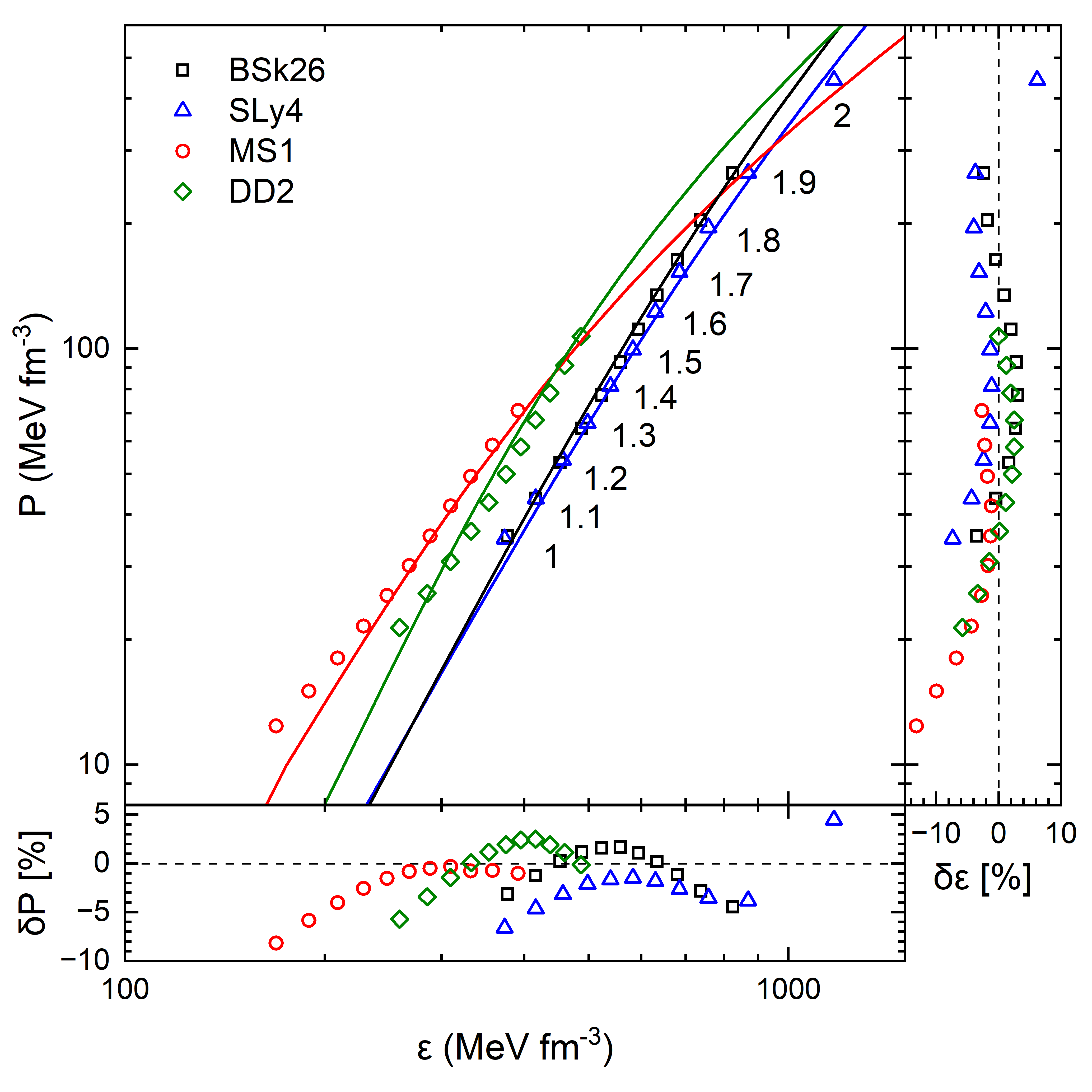}
    \includegraphics[width=9cm]{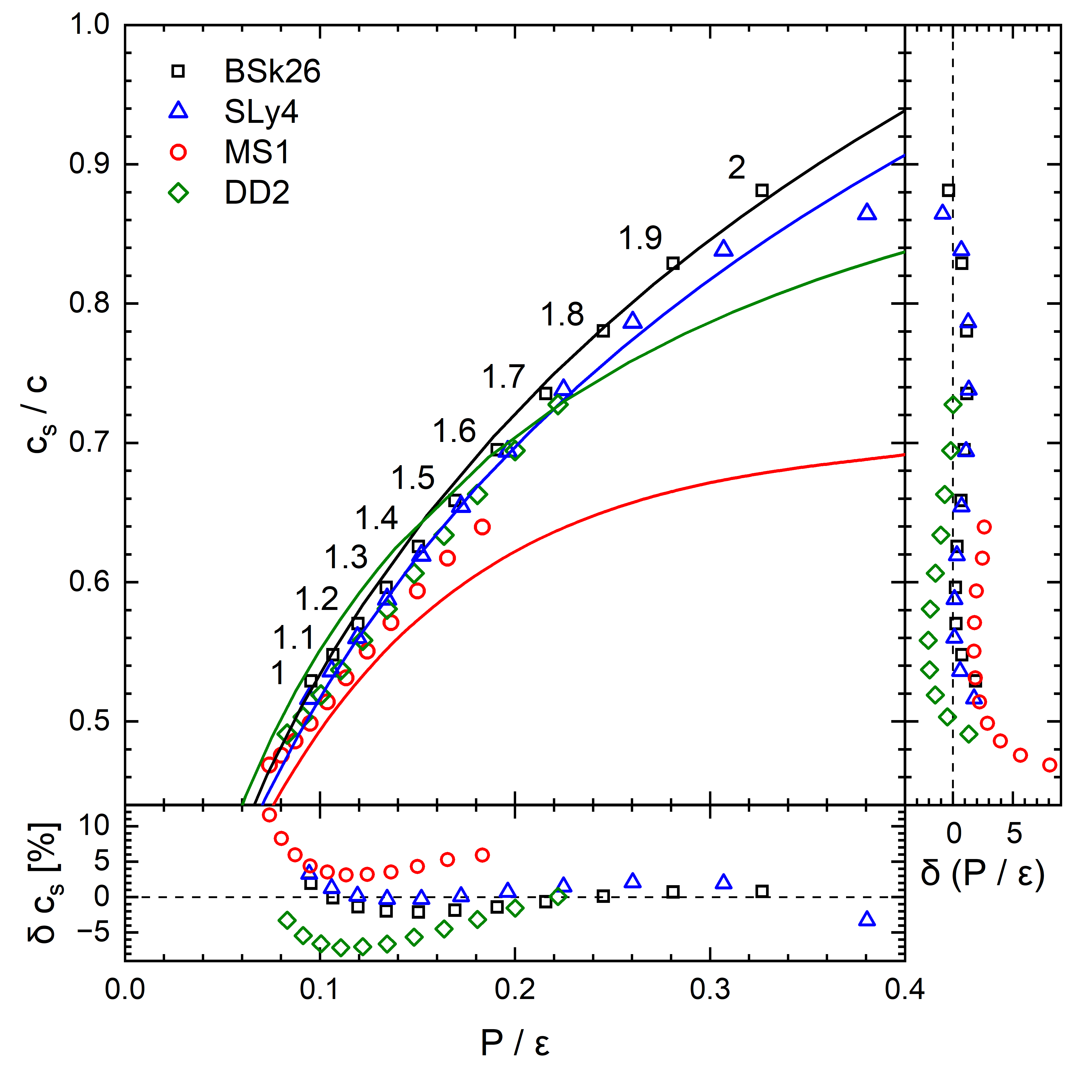}
    \caption{Comparison of two-radius and inverse slope power-law fits for the general mass-radius inversion technique. Left panel:  Solid lines show the $P\mbox{-}\mathcal{E}$ relation for the BSk26, SLy4, MS1 and DD2 equations of state.  Points show the reconstructed $\mathcal{E}_c$ and $P_c$ from Eq. (\ref{eq:powfit4}) at the indicated reference masses.  The side and lower panels show their logarithmic errors at each mass point.  Right panel: Solid lines show the $c_s/c\mbox{-}P/\mathcal{E}$ relation, while points show the reconstructed values of $c_{s,c}/c$ and $\mathcal{E}_c/P_c$ from Eq. (\ref{eq:G2})  at their indicated reference masses.  Side and lower panels show their logarithmic errors.
    }
\label{fig:2Rcomp1}\end{figure}

\begin{figure}[h]
\vspace*{-1.6cm}
\hspace*{-1.35cm}
\includegraphics[width=21cm,angle=180]{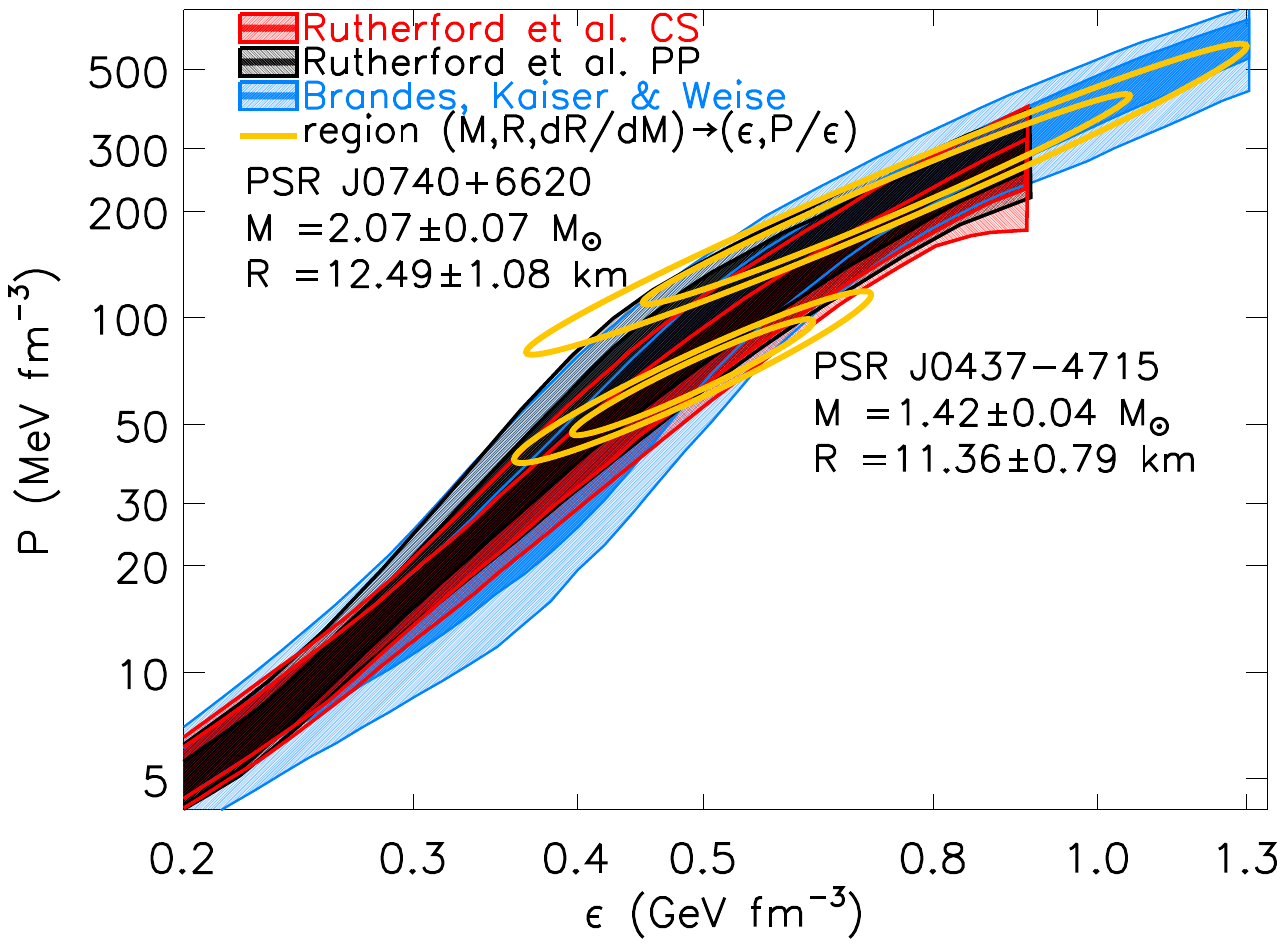}

\vspace*{-1.4cm}
\caption{The lower and upper gold confidence ellipses (68\% and 95\%) correspond to the inversion of the corresponding PSR J0437-4715 and J0740+6620 $M\mbox{-}R$ uncertainty regions, respectively, both assumed to be uncorrelated double Gaussian probability distributions, using Eq. (\ref{eq:powfit5}) for $\mathcal{E}$ and $\hat P=P/\mathcal{E}$ and their uncertainties (see text for determination of inverse slopes $dR/dM$).  For comparison, results for the inferred $\mathcal{E}\mbox{-}P$ relations from traditional Bayesian approaches of  \cite{Brandes2023} (blue) and \cite{Rutherford2024} for their sound-speed CS (piecewise polytrope PP) parameterization are shown in red (black). 68\% (95\%) bounds are shown with bolder (lighter) shades.}
\label{fig:PE-MRdRdM}
\end{figure}
An example of how this approach could be applied to give EOS information from observations is shown in Fig. \ref{fig:PE-MRdRdM}.  For this example, we utilized the recent NICER results for two sources, PSR J0437-4715 \citep{Dittman2024,Salmi2024}  and J0740+6620 \citep{Choudhury2024}, whose masses, about $1.4M_\odot$ \citep{Fonseca2021} and $2.0M_\odot$ \citep{Reardon2024}, respectively, were independently established from pulsar timing.  The assumed uncertainty regions were approximated by double uncorrelated Gaussian probability distributions with $M=1.42\pm0.04M_\odot$ and $R=11.36\pm0.79$ km for PSR J0437-4715 and $M=2.07\pm0.07M_\odot$ and $R=12.49\pm1.08$ km for PSR J0740+6620.  Randomly sampling from the two regions, one can approximate values of $dR/dM$ from finite differencing for each pair of points from the two regions.  Including the additional RMS uncertainties of the fits provided by Eq. (\ref{eq:powfit5}) as given in Table \ref{tab:genfit1}, the observational uncertainty regions can be transformed into the two sets of EOS confidence ellipses in Fig. \ref{fig:PE-MRdRdM}.  Each set of confidence ellipses shows a strong correlation between $\mathcal{E}$ and $P$ because the formulae Eq. (\ref{eq:powfit5}) have much smaller uncertainties for $\hat P$ than for either $\mathcal{E}$ or $P$.

Of course, one cannot reconstruct an entire $\mathcal{E}\mbox{-}P$ band from this limited observational information.  Traditional Bayesian approaches circumvent this difficulty by creating prior distributions of $M\mbox{-}R$ relations from parameterized EOS models, statistically determining those $M\mbox{-}R$ curves that best-fit the observations, and then recovering probability distributions for $\mathcal{E}\mbox{-}P$ that result in a continuous band.  We compare the results of our approach with those of traditional Bayesian methods in Fig. \ref{fig:PE-MRdRdM} for three published cases in which the dominant observational information that was utilized were from the same two NICER sources.  The comparisons are with \cite{Brandes2023}, who parameterized the EOS with a variable sound-speed approach, and with \cite{Rutherford2024} who parameterized the EOS with first, a piecewise-polytrope (PP) model, and second, with a piecewise constant sound-speed (CS) model.

This figure illustrates the prior uncertainties introduced by choices of EOS parameterization and/or parameter sampling, which are more important at lower densities ($2.5n_s$ is equivalent to $\mathcal{E}\sim0.4{\rm MeV~fm}^{-3}$) than higher densities ($5n_s$ is equivalent to $\mathcal{E}\sim0.7{\rm MeV~fm}^{-3}$) near the central density of the maximum mass star.  As shown by \cite{LP01}, radii of typical neutron stars are determined by the EOS near $2n_s$ and not near the central densities of maximum mass stars, so the differences shown in Fig. \ref{fig:PE-MRdRdM} have implications for predicted radii.  The three different Bayesian approaches have similar uncertainties, but the centroids of their EOS results differ by about $0.25\sigma$ at high densities and more than $0.5\sigma$ at lower densities.  In other words, the uncertainties stemming from the parameterization choice can be nearly as large as those from observational uncertainties, especially for lower mass stars.
Within this context,  the predicted sound speed of our approach appears to be slightly more consistent with the results of \cite{Brandes2023} than \cite{Rutherford2024} which suggest a slightly lower sound speed around $\mathcal{E}\simeq0.5$ GeV fm$^{-3}$.

\section{Discussion and Conclusion\label{sec:discuss}}

In this paper, we developed fitting formulae for the central values of $\mathcal{E}, P, c_s,\mu$ and $n$ for masses given by specific fractions of the maximum mass.  These formulae were optimized for hadronic EOSs and are EOS-insensitive with accuracies of 0.5\% or better.  The formulae can be used to invert a specific $M\mbox{-}R$ curve to yield its underlying EOS ($\mathcal{E}\mbox{-}P$ relation).  The question arises whether they can also be applied to $M\mbox{-}R$ curves for stars with first-order phase transitions.

While we did not perform an exhaustive study, the case of a typical hadronic EOS combined with the simple MIT massless, two-flavor, charge-neutral quark bag model is illuminating.  For this exercise, we assumed a quark bag EOS given by
\begin{equation}
    \mathcal{E}_{bag}=3P_{bag}+4B={3\over4}\mu_{bag} n_{bag}+B,
    \label{eq:bag}
    \end{equation}
where $B=80$ MeV fm$^{-3}$ was chosen for the bag constant.   $\mu_{bag}$ and $n_{bag}$ are
the baryon chemical potential and baryon number density in the quark phase.  For each quark species, the Fermi gas model for degenerate, massless up ($u$) and down ($d$) quarks gives
\begin{equation}
\mu_{u,d}=\hbar c\left({3\over N_cg}\pi^2n_{u,d}\right)^{1/3},
\label{eq:bag1}
\end{equation}
where $g=2$ is the spin degeneracy and $N_c=3$ is the number of quark colors.  Charge neutrality results in $n_d=2n_u$ or $\mu_d=2^{1/3}\mu_u$, and therefore $n_{bag}=N_cn_d/2$.  The baryon chemical potential is then related to the baryon density by
\begin{equation}
    \mu_{bag}=2\mu_d+\mu_u=\hbar c\left(2^{4/3}+1\right)\left(\pi^2n_{bag}\right)^{1/3}.
\end{equation}
At zero pressure, one has $\mathcal{E}_0=4B=\mu_0n_0$, where the chemical potential and number densities are
\begin{equation}
    \mu_0 =\left[\hbar c\left(2^{4/3}+1\right)\right]^{3/4}\pi^{1/2}(4B)^{1/4}, \quad n_0 = \frac{4B}{\mu_0}.
\end{equation}
With $B=80$ MeV fm$^{-3}$, $\mu_0=1.014$ GeV and $n_0=0.3155$ fm$^{-3}$.  With these quantities, one can express the baryon density and pressure more simply as
\begin{equation}
    n_{bag} = n_0\left(\frac{P_{bag}+B}{B}\right)^{3\over4}, \quad P_{bag} = B\left[\left(\frac{n_{bag}}{n_0}\right)^{4\over3}-1\right].
\end{equation}
 Paired with a hadronic EOS at low densities, a first order phase transition to quark matter is found from the conditions $\mu_{bag}=\mu_h$ and $P_{bag}=P_h$, where the subscript h refers to the hadronic EOS (BSk22 in this case).  With $B=80$ MeV fm$^{-3}$, the phase transition occurs at the pressure $P_t=31.92$ MeV fm$^{-3}$ and chemical potential $\mu_t=1.103$ GeV, with a baryon density discontinuity extending from $n_{ht}=0.304$ fm$^{-3}$ to $n_{qt}=0.406$ fm$^{-3}$, and an energy density discontinuity extending from $\mathcal{E}_{ht}=303.8$ MeV fm$^{-3}$ to $\mathcal{E}_{qt}=415.8$ MeV fm$^{-3}$.  This transition has a discontinuity magnitude $\mathcal{E}_{qt}/\mathcal{E}_{ht}-1=0.37$, close to the largest seemingly allowed by neutron star maximum mass and measured radius constraints \citep{Brandes2023}.

\begin{figure}[h]
\hspace*{-0.9cm}    \includegraphics[width=10.2cm,angle=180]{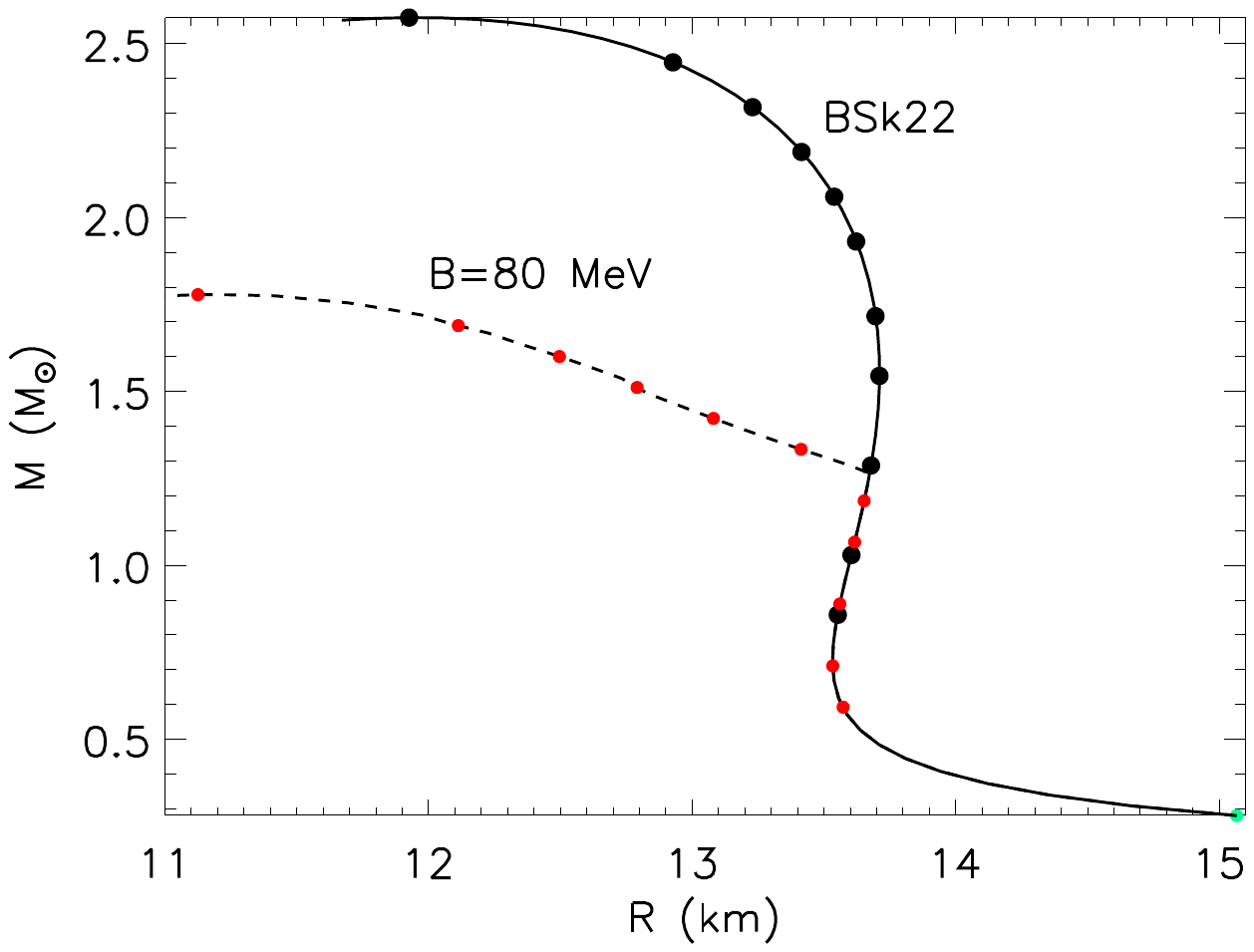}
\hspace*{-2cm}    \includegraphics[width=10.2cm,angle=180]{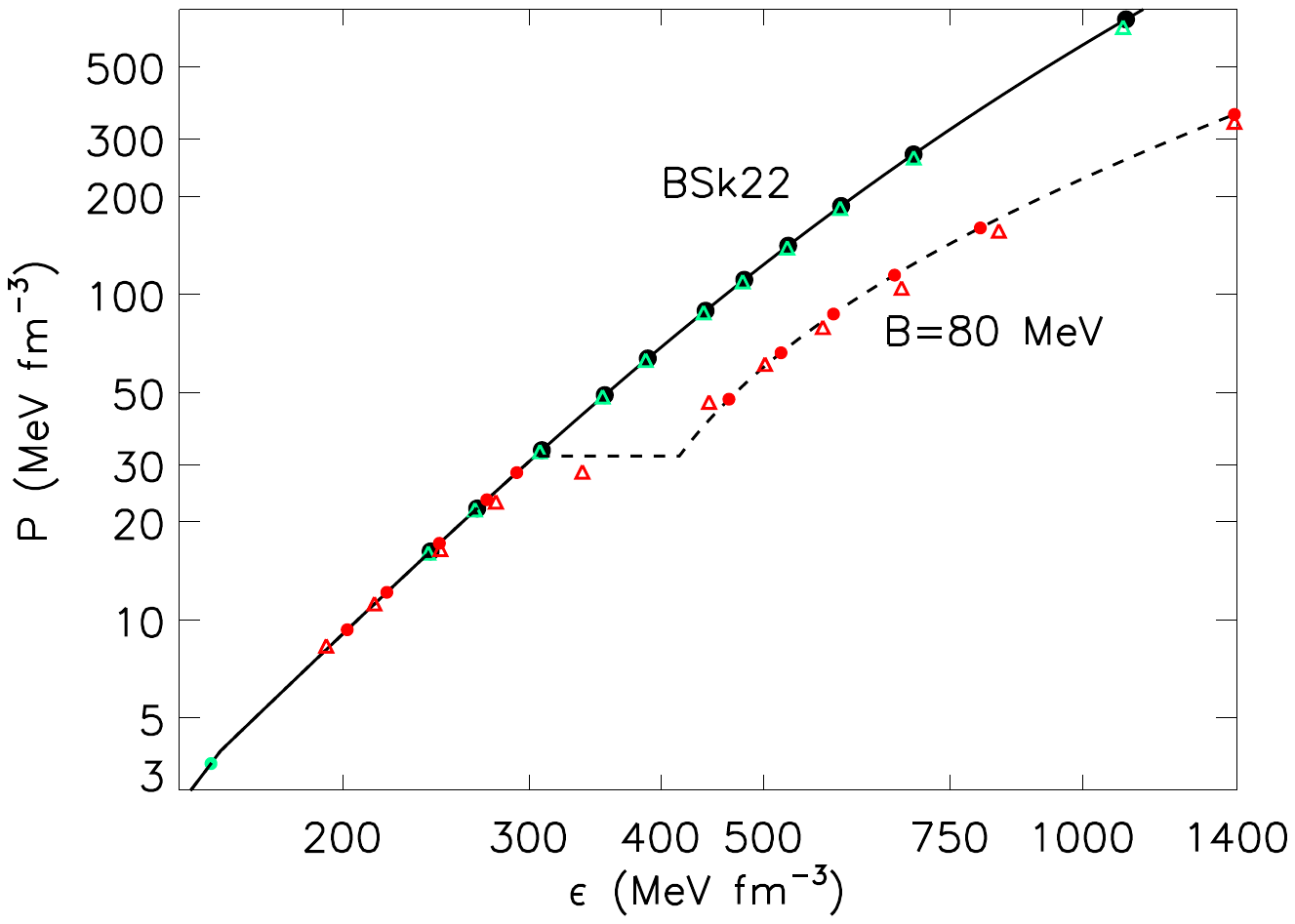}
 \vspace*{-1cm}
 \caption{Left panel: mass-radius curves for the purely hadronic star formed with the BSk22 EOS (solid curve) and a hybrid star with a first-order phase transition to the MIT massless, charge neutral, two-flavor quark bag model EOS with $B=80$ MeV (dashed curve).  Black (red) filled circles show the fractional maximum mass ($M_f,R_f$) values.  Right panel: $\mathcal{E}\mbox{-}P$ relations forming the $M\mbox{-}R$ diagrams shown in the left panel.  Black (red) filled circles show the actual central values for $\mathcal{E}_c$ and $P_c$ for the BSK22 (BSK22+Bag) case.  The black (red) triangles show the predicted values for $\mathcal{E}_c$ and $P_c$ from Eq. (\ref{eq:eqfit1}) using parameters from Table \ref{tab:powfit1}.  Green filled circles in both panels show properties at the nuclear saturation density $n_s$.}
\label{fig:BSk22-bag}\end{figure}

\begin{figure}
\centering
%\hspace*{-0.9cm}    
\includegraphics[width=10.2cm,angle=180]{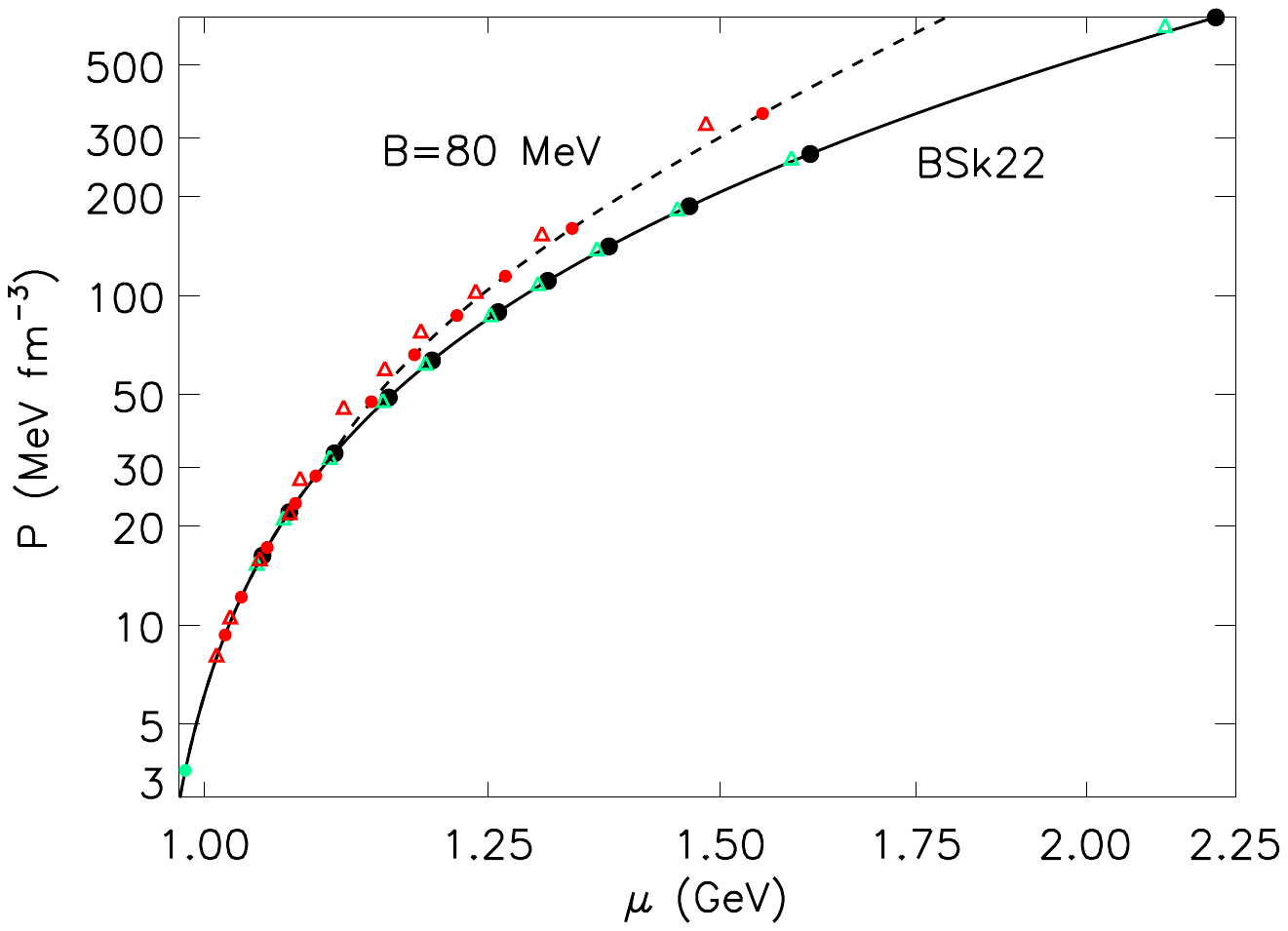}
%\hspace*{-2cm}    \includegraphics[width=10.2cm,angle=180]{PE-BSK22Bag.pdf}
 \vspace*{-1cm}
 \caption{The same as the right panel of Fig. \ref{fig:BSk22-bag} except for the pressure and chemical potential.}
\label{fig:BSk22-bag1}\end{figure}

The mass-radius curves with and without this phase transition are displayed in Fig. \ref{fig:BSk22-bag}, showing the maximum mass is reduced from $2.58M_\odot$ to $1.78M_\odot$.  This figure also displays the underlying EOSs for the two cases.  The interpolated  central values $\mathcal{E}_c$ and $P_c$ at the fractional maximum masses $M_f$ are compared with predictions stemming from Eq. (\ref{eq:eqfit1}) using the parameters specified in Table \ref{tab:powfit1}.  In Fig. \ref{fig:BSk22-bag1} the corresponding interpolated central values $\mu_c$ and $P_c$ are compared with predictions. 

As expected, the reconstruction of the BSk22 EOS from its $M\mbox{-}R$ curve is highly accurate (to within 0.5\%) for all fractional maximum masses.  The reconstruction of the hybrid case with the first-order phase transition is obviously not as precise, although it is by no means disastrous.  Since the maximum masses are different for the two cases, the actual $\mathcal{E}_c$, $\mu_c$ and $P_c$ values at each $M_f$ are different.  Generally speaking, the reconstructed values of $\mathcal{E}_c, \mu_c$ and $P_c$ far from the phase transition are relatively well reproduced, with accuracies of a few percent.  Near the phase transition, the reconstructed EOS fails to reproduce the abrupt behavior of the actual $\mathcal{E}_c\mbox{-}P_c$ relation, but instead smooths out the transition.  Nevertheless, at the energy-density midpoint of the transition, the central pressure and energy density are accurately reproduced.  The reconstructed pressures and chemical potentials in the quark phase are both somewhat underestimated, but the predicted values still lie near the interpolated EOS curve.  Overall, the EOS reconstruction is surprisingly satisfactory, indicating that this method can be applied to dense matter EOSs rather different from the hadronic Skyrme and RMF forces used to establish the fitting formulae, even those including large first-order phase transitions.  The expectation is our formulae will also be applicable to other hybrid star models, such as those with crossover transitions involving quarkyonic matter \citep{McLerran2019,Zhao2020}.

Given the relative accuracies of the reconstructed EOSs using our formulae, and the large existing observational uncertainties, it would not seem essential to further refine the techniques developed here to obtain higher accuracies.   If needed, however, two obvious ways of increasing the inversion precision would be to select a finer fractional maximum mass grid with more values of $f$, or to select more than two fitting radii at each fractional maximum mass value. We have not yet explored how the accuracy depends on either of these considerations.  Nevertheless, there may be a fundamental limit to precision based on the training sample of EOSs that we have employed.  A future project will be to utilize several versions of parameterized EOSs to validate the fitting parameters found here.

We have found, in any case, that one cannot simply perform Newton-Raphson iterations to refine the predicted values of $\mathcal{E}_c$ and $P_c$ from $M_f-R_f$ values using the TOV equation.  Such iterations are highly unstable, which is what motivated \cite{Lindblom92} to develop a more complicated method to estimate the EOS from the mass-radius curve.  As we have noted, Lindblom's technique still has general instabilities which reduce its utility.  A project for future work will be to refine our predictions, possibly using a variation of Lindblom's technique together with predicted sound speed information, which would be most useful for applications to cases with first-order phase transitions.  But given  observational uncertainties, such refinements would not seem to be crucial at the present time.

Neutron star masses and radii are not the only observable structural properties.  Our methods should also work for moments of inertia ($\bar I$) and tidal deformabilities ($\Lambda$), which can be shown to have strong correlations with $M$ and $R$ \citep{Zhao2018}.  In principle, power-law formulae can be constructed for the inference of $\mathcal{E}_c, P_c, \mu_c$ and $n_c$ directly from the maximum mass and either of the dimensionless quantities $\Lambda_f$ or $\bar I_f$.  The fact that $\Lambda$ and $\bar I$ are highly correlated to better than the 1\% level \citep{Yagi2013} irrespective of EOS or mass means that they could be simultaneously utilized.  

Future work will also be directed toward combining our formulae with traditional Bayesian techniques for inferring the EOS from observations.  One possibility is to directly parameterize $M-R$ relations rather than to generate them from parameterized EOSs.  One could then attempt to assign probabilities to the $M\mbox{-}R$ relations that reflect their overlap  with the observational uncertainty regions. Our formulae can then be used to generate an uncertainty band in $\mathcal{E}_P$ space from the resulting probability distributions of $M\mbox{-}R$ curves, while avoiding the problem of prior uncertainties stemming from EOS parameterizations. 
\\
\\
%\begin{acknowledgments}
We acknowledge funding from the US Department of Energy under Grant DE-FG02-87ER40317.  The seeds for this research were planted at the The Modern Physics of Compact Stars and Relativistic Gravity 2023 Conference in Yerevan, Armenia.  Discussions with Dima Ofengeim, Armen Sedrakian and Sophia Han are very much appreciated.  We thank Wolfram Weise and Len Brandes  for details of their Bayesian analyses.
%\end{acknowledgments}

%% To help institutions obtain information on the effectiveness of their 
%% telescopes the AAS Journals has created a group of keywords for telescope 
%% facilities.
%
%% Following the acknowledgments section, use the following syntax and the
%% \facility{} or \facilities{} macros to list the keywords of facilities used 
%% in the research for the paper.  Each keyword is check against the master 
%% list during copy editing.  Individual instruments can be provided in 
%% parentheses, after the keyword, but they are not verified.

\bibliography{correlation}

\begin{thebibliography}{130}
\expandafter\ifx\csname natexlab\endcsname\relax\def\natexlab#1{#1}\fi


\bibitem[{{Zhao} \& {Lattimer}(2020)}]{2020PhRvD.102b3021Z}
{Zhao}, T. \& {Lattimer}, J.~M. 2020, \prd, 102, 023021
\end{thebibliography}

\begin{thebibliography}{}
\expandafter\ifx\csname natexlab\endcsname\relax\def\natexlab#1{#1}\fi
\providecommand{\url}[1]{\href{#1}{#1}}
\providecommand{\dodoi}[1]{doi:~\href{http://doi.org/#1}{\nolinkurl{#1}}}
\providecommand{\doeprint}[1]{\href{http://ascl.net/#1}{\nolinkurl{http://ascl.net/#1}}}
\providecommand{\doarXiv}[1]{\href{https://arxiv.org/abs/#1}{\nolinkurl{https://arxiv.org/abs/#1}}}

\bibitem[{Brandes {et~al.}(2023)Brandes, Weise, \& Kaiser}]{Brandes2023}
Brandes, L., Weise, W., \& Kaiser, N. 2023, Phys. Rev. D, 108, 094014, \dodoi{10.1103/PhysRevD.108.094014}

\bibitem[{Cai {et~al.}(2023{\natexlab{a}})Cai, Li, \& Zhang}]{Cai2023}
Cai, B.-J., Li, B.-A., \& Zhang, Z. 2023{\natexlab{a}}, The Astrophysical Journal, 952, 147, \dodoi{10.3847/1538-4357/acdef0}

\bibitem[{Cai {et~al.}(2023{\natexlab{b}})Cai, Li, \& Zhang}]{Cai2023vs}
---. 2023{\natexlab{b}}, Phys. Rev. D, 108, 103041, \dodoi{10.1103/PhysRevD.108.103041}

\bibitem[{Chabanat {et~al.}(1998)Chabanat, Bonche, Haensel, Meyer, \& Schaeffer}]{chabanat1998skyrme}
Chabanat, E., Bonche, P., Haensel, P., Meyer, J., \& Schaeffer, R. 1998, Nucl. Phys. A, 635, 231

\bibitem[{Choudhury {et~al.}(2024)Choudhury, Salmi, Vinciguerra, \& et~al.}]{Choudhury2024}
Choudhury, D., Salmi, T., Vinciguerra, S., \& et~al. 2024, Ap. J. Lett., 971, L20

\bibitem[{Dittmann {et~al.}(2024)Dittmann, Miller, Lamb, \& et~al.}]{Dittman2024}
Dittmann, A.~J., Miller, M.~C., Lamb, F.~K., \& et~al. 2024, Ap. J., 974, 295

\bibitem[{Fonseca {et~al.}(2021)Fonseca, Cromartie, Pennucci, \& et~al.}]{Fonseca2021}
Fonseca, E., Cromartie, H.~T., Pennucci, T.~T., \& et~al. 2021, Ap. J. Lett., 915, L12

\bibitem[{Grinstead \& Snell(1997)}]{Grinstead1997}
Grinstead, C.~M., \& Snell, J.~L. 1997, Introduction to Probability (Providence, RI: American Mathematical Society)

\bibitem[{{Lattimer} \& {Prakash}(2001)}]{LP01}
{Lattimer}, J.~M., \& {Prakash}, M. 2001, \apj, 550, 426, \dodoi{10.1086/319702}

\bibitem[{Lattimer \& Prakash(2011)}]{Lattimer2011}
Lattimer, J.~M., \& Prakash, M. 2011, What a Two Solar Mass Star Really Means, ed. S.~Lee (Singapore: World Scientific)

\bibitem[{Lindblom(1992)}]{Lindblom92}
Lindblom, L. 1992, Astrophys. J., 398, 569

\bibitem[{McLerran \& Reddy(2019)}]{McLerran2019}
McLerran, L., \& Reddy, S. 2019, Phys Rev Lett, 122, 122701

\bibitem[{Ofengeim(2020)}]{Ofengeim2020}
Ofengeim, D.~D. 2020, Phys. Rev. D, 101, 103029

\bibitem[{Ofengeim {et~al.}(2023)Ofengeim, Shternin, \& Piran}]{Ofengeim2023}
Ofengeim, D.~D., Shternin, P.~S., \& Piran, T. 2023, Astronomy Letters, 49, 567–574, \dodoi{10.1134/s1063773723100055}

\bibitem[{Oppenheimer \& Volkoff(1939)}]{Oppenheimer39}
Oppenheimer, J.~R., \& Volkoff, G.~M. 1939, Phys. Rev., 55, 374

\bibitem[{Reardon {et~al.}(2024)Reardon, Bailes, Flynn, \& et~al.}]{Reardon2024}
Reardon, D.~J., Bailes, M., Flynn, C., \& et~al. 2024, Ap. J. Lett., 971, L18

\bibitem[{Rhoades \& Ruffini(1974)}]{Rhoades1974}
Rhoades, C.~E., \& Ruffini, R. 1974, Phys. Rev. Lett., 32, 324

\bibitem[{Rutherford {et~al.}(2024)Rutherford, Mendes, Svensson, Schwenk, Watts, Hebeler, Keller, Prescod-Weinstein, Choudhury, Raaijmakers, Salmi, Timmerman, Vinciguerra, Guillot, \& Lattimer}]{Rutherford2024}
Rutherford, N., Mendes, M., Svensson, I., {et~al.} 2024, The Astrophysical Journal Letters, 971, L19, \dodoi{10.3847/2041-8213/ad5f02}

\bibitem[{Salmi {et~al.}(2024)Salmi, Choudhury, Kini, \& et~al.}]{Salmi2024}
Salmi, T., Choudhury, D., Kini, Y., \& et~al. 2024, Ap. J., 974, 294

\bibitem[{Sun {et~al.}(2024{\natexlab{a}})Sun, Bhattiprolu, \& Lattimer}]{Sun2024}
Sun, B., Bhattiprolu, S., \& Lattimer, J.~M. 2024{\natexlab{a}}, Phys. Rev. C, 109, 055801, \dodoi{10.1103/PhysRevC.109.055801}

\bibitem[{Sun {et~al.}(2024{\natexlab{b}})Sun, Xu, \& Lattimer}]{zenodo_TOV}
Sun, B., Xu, K., \& Lattimer, J. 2024{\natexlab{b}}, Visualization Software for Analytic Inversion of an Arbitrary Neutron Star M-R Curve into its Underlying Pressure-Energy Density Relation,  Zenodo, \dodoi{10.5281/zenodo.14064108}

\bibitem[{Tolman(1934)}]{Tolman34}
Tolman, R.~C. 1934, Relativity, Thermodynamics and Cosmology (Oxford: Clarendon Press)

\bibitem[{Yagi \& Yunes(2013)}]{Yagi2013}
Yagi, K., \& Yunes, N. 2013, Science, 341, 365

\bibitem[{Zhao \& Lattimer(2018)}]{Zhao2018}
Zhao, T., \& Lattimer, J.~M. 2018, Phys Rev D, 98, 063030

\bibitem[{Zhao \& Lattimer(2020)}]{Zhao2020}
---. 2020, Phys Rev D, 102, 023021

\bibitem[{Zhao \& Lattimer(2022)}]{Zhao2022}
---. 2022, Phys. Rev. D, 106, 123002, \dodoi{10.1103/PhysRevD.106.123002}

\end{thebibliography}
%\bibliographystyle{aasjournal}

%% This command is needed to show the entire author+affiliation list when
%% the collaboration and author truncation commands are used.  It has to
%% go at the end of the manuscript.
%\allauthors

%% Include this line if you are using the \added, \replaced, \deleted
%% commands to see a summary list of all changes at the end of the article.
%\listofchanges

\end{document}